\documentclass[acmsmall]{acmart}
\usepackage{amsmath,mathtools}
\usepackage[ruled,vlined]{algorithm2e}
\usepackage{multirow}
\usepackage{graphicx}
\usepackage{subfigure}
\usepackage{caption}
\usepackage{booktabs}
\usepackage{xcolor,colortbl}

\AtBeginDocument{%
  \providecommand\BibTeX{{%
    \normalfont B\kern-0.5em{\scshape i\kern-0.25em b}\kern-0.8em\TeX}}}

\setcopyright{acmlicensed}
\acmJournal{PACMHCI}
\acmYear{2022} \acmVolume{6} \acmNumber{CSCW2} \acmArticle{371} \acmMonth{11} \acmPrice{15.00}\acmDOI{10.1145/3555096}

\newcommand\finaledit[1]{{\color{black} #1}}

\newcommand\jedit[1]{{\color{black} #1}}
\newcommand\anedit[1]{{\color{black} #1}}
\newcommand\aledit[1]{{\color{black} #1}}

\begin{document}

%% \title{Understanding Effects of Algorithmic vs. \finaledit{Community} Label on Perceived Accuracy of Hyper-partisan Misinformation}

\title[Understanding Effects of Algorithmic vs. Community Label on Perceived Accuracy \\ of Hyper-partisan Misinformation]{Understanding Effects of Algorithmic vs. Community Label on Perceived Accuracy of Hyper-partisan Misinformation}


\author{Chenyan Jia}
\email{chenyanjia@utexas.edu}
% \orcid{0000-0002-8407-9224}
\affiliation{%
  \institution{Moody College of Communication, The University of Texas at Austin}
  \streetaddress{300 W. Dean Keeton, A0900}
  \city{Austin}
  \state{Texas}
  \country{USA}
  \postcode{78712–1069}
}

\author{Alexander Boltz}
\email{alexboltz@utexas.edu}
\authornotemark[1]
\affiliation{%
  \institution{Department of Government, The University of Texas at Austin}
%   \streetaddress{116 Inner Campus Dr Stop G6000}
%   \city{Austin}
 \country{USA}
%   \postcode{78712}
}

\author{Angie Zhang}
\email{angie.zhang@austin.utexas.edu}
\authornote{Both authors contributed equally to this research.}
\affiliation{%
  \institution{School of Information, The University of Texas at Austin}
%   \streetaddress{1616 Guadalupe St, Suite #5.202}
%   \city{Austin}
\country{USA}
%   \postcode{78701-1213}
}

\author{Anqing Chen}
\email{benjamin.c0427@gmail.com}
\affiliation{%
  \institution{Department of Electrical and Computer Engineering, The University of Texas at Austin}
%   \streetaddress{2501 Speedway, C0803}
%   \city{Austin}
\country{USA}
%   \postcode{78712}
}

\author{Min Kyung Lee}
\email{minkyung.lee@utexas.edu}
\affiliation{%
  \institution{School of Information, The University of Texas at Austin}
%   \streetaddress{1616 Guadalupe St, Suite #5.202}
%   \city{Austin}
 \country{USA}
%   \postcode{78701-1213}
}

\renewcommand{\shortauthors}{Chenyan Jia et al.}
%% If needed use a foot or author note to identify equal contribution

\begin{abstract}
  Hyper-partisan misinformation has become a major public concern. In order to examine what type of \finaledit{misinformation} label can mitigate hyper-partisan misinformation sharing on social media, we conducted a 4 (label \finaledit{type}: algorithm, \finaledit{community}, third-party fact-checker, \finaledit{and} no label) X 2 (\finaledit{post} ideology: liberal vs. conservative) between-subjects online experiment ($N$ = 1,677) in the context of COVID-19 \finaledit{health information}. \finaledit{The results suggest that for liberal users, all labels reduced the perceived accuracy and believability of fake posts regardless of the posts' ideology. In contrast, for conservative users, the efficacy of the labels depended on whether the posts were ideologically consistent: algorithmic labels were more effective in reducing the perceived accuracy and believability of fake conservative posts compared to community labels, whereas all labels were effective in reducing their belief in liberal posts.} Our results shed light on the differing effects of various misinformation labels \finaledit{dependent} on \finaledit{people's political ideology}.
\end{abstract}


%   <concept>
%       <concept_id>10003120.10003130.10011762</concept_id>
%       <concept_desc>Human-centered computing~Empirical studies in collaborative and social computing</concept_desc>
%       <concept_significance>500</concept_significance>
%       </concept>
%  </ccs2012>
% \ccsdesc[500]{Human-centered computing~Empirical studies in collaborative and social computing}

\begin{CCSXML}
<ccs2012>
   <concept>
       <concept_id>10003120.10003121.10011748</concept_id>
       <concept_desc>Human-centered computing~Empirical studies in HCI</concept_desc>
       <concept_significance>500</concept_significance>
       </concept>
 </ccs2012>
\end{CCSXML}

\ccsdesc[500]{Human-centered computing~Empirical studies in HCI}

\keywords{accuracy, misinformation label, hyper-partisan misinformation, social media}

\maketitle

\section{Introduction}

Hyper-partisan misinformation has become a major public concern that can exacerbate partisan disagreement over even basic facts \cite{lazer2018science, pennycook2021shifting}. Partisans tend to believe news that aligns with their beliefs regardless of truthfulness due to confirmation bias \cite{kim2019combating}. Such confirmation bias has proven to be one major factor affecting social media users’ belief in news articles \cite{kim2019combating}. Partisans may share ideologically consistent but false information on social media, partially because their attention focuses more on the political alignment of content instead of accuracy \cite{moravec2020appealing, van2018partisan}. Previous studies have identified partisanship as a stronger predictor of misinformation sharing than veracity among Twitter users \cite{grinberg2019fake}.

The widespread sharing of hyper-partisan misinformation on social media is problematic and can be connected to a couple factors. \jedit{First, news consumption on social media is often less mindful because of the entertainment-seeking goals of social media use \cite{johnson2015reasons, kim2019says}. Frequent social media users often process information with hedonic mindsets \cite{van2017spontaneous} and thus are less likely to think critically than when they are in a utilitarian mindset \cite{thatcher2018mindfulness}}. Additionally, social media platforms further accelerate the proliferation of false and misleading information because of numerous sharing features of the platforms and the sheer number of users \cite{epstein2020will, silverman2016analysis}. Researchers have expressed their fears of the potential for social media platforms being leveraged by propagandists with ulterior motives such as confusing voters with information overload, preventing them from being able to distinguish truths from falsehoods \cite{faris2017partisanship, liu2021transformer}. One recent study suggests that prior exposure to fake political posts, even for extremely implausible posts, can increase people's perceived accuracy of such misinformation \cite{pennycook2018prior}.

Researchers and practitioners have recognized the urgent need for seeking solutions to combat the spread of misinformation. Many previous studies have shown that misinformation labels (e.g., stop signs, disclaimers, or warnings) can reduce people's believability of fake posts and decrease the propensity to share misinformation \cite{moravec2020appealing, yaqub2020effects}. Most past studies have examined the effect of labels attributed to third-party fact-checkers \cite{moravec2020appealing, pennycook2020implied} or human moderators hired by social media platforms. \jedit{For instance, previous work examined the impact of third-party fact-checkers' labels on Twitter or Facebook posts containing false election claims during the 2020 Presidential election \cite{zannettou2021won}}. Although those labels provide possible solutions to mitigate misinformation sharing, this approach to labelling has a pressing limitation: in spite of its high credibility and reliability, third-party fact-checkers' labels heavily rely on human moderation and can not provide real-time intervention \cite{yaqub2020effects}. Thus by the time posts get manually labelled, the false information may have already spread. 

One alternative approach is to leverage real-time interventions such as automated labelling (i.e., through misinformation detection algorithms) or community-based methods to assist users in distinguishing false information from factual content. 
While the development of misinformation detection algorithms continues to expand \cite{jain2016towards,horne2017just,hosseinimotlagh2018unsupervised} and Twitter has launched a preliminary pilot (Birdwatch\footnote{blog.twitter.com/en\_us/topics/product/2021/introducing-birdwatch-a-community-based-approach-to-misinformation.html}) to test crowdsourced misinformation reporting, not much additional work has yet explored how to use these techniques to provide real-time feedback to users. Very few studies have examined how artificial intelligence (AI) and human-related labels differ in reducing people’s believability in misinformation. One pioneering paper explores how the presence of an algorithmic misinformation detection warning and a fact-checker's warning will influence people’s ability to detect misinformation \cite{seo2019trust}. Results show that those participants made more correct decisions identifying misinformation with the algorithmic warning (78.3\%) than without any warning (70.1\%), but their accuracy in identification of real news was similar between the two conditions \cite{seo2019trust}. Another work examined the impact of real-time labels such as misinformation detection algorithms and other users from the community \cite{yaqub2020effects}. Results showed that both the algorithmic label and community users' label succeeded in persuading people to avoid sharing false headlines, but both were less effective compared to third-party labels. Neither of the papers, however, measured the participants’ political stance as a factor and thus overlooked whether there may have been a difference in misinformation labels' impact on posts \finaledit{dependent on people's political ideology}. To address this gap, our paper intends to further investigate the different effects of real-time labels (i.e., labels attributed to algorithms and \finaledit{community}) and third-party fact-checkers' labels on partisans' perceptions of hyper-partisan social media posts. \jedit{Since 2020, the COVID-19 pandemic has posed an unprecedented risk to society, resulting in a polarized political landscape \cite{pennycook2020predictors}. Prior work found initial evidence that people's misperceptions about COVID-19 are associated with people's political ideology \cite{pennycook2020predictors}. Given the emerging nature of such a topic, our paper focuses on COVID-19 related social media posts.} 

In recent years, there has been a prevalent expectation that AI will have a unique advantage in online political contexts \cite{wojcieszak2021can}. Public discourse reflects that AI is often perceived as more fair, objective, unbiased, and as having a less political agenda under the assumption of neutrality \cite{gillespie2014relevance}. People tend to hold positive stereotypes about machine infallibility and neutrality \cite{sundar2019machine}. Such heuristics may generate an advantage for AI-related products in political settings, particularly where these products may assist in mitigating hyper-partisan misinformation. Previous studies have examined AI-related decisions vs. human decisions in the context of news recommendation and online content moderation \cite{wojcieszak2021can, wang2021moderating}. Initial evidence suggests that people’s perceptions of news bias may be attenuated when news is attributed to a machine cue \cite{wang2021moderating} because people are inclined to believe that stories written or selected by a machine must be objective and free from political bias \cite{waddell2019can, jia2021source}.

Nevertheless, not many studies have examined algorithmic labels vs. human labels in the context of hyper-partisan misinformation sharing on social media. One pioneering work finds no difference between the effect of algorithmic and other users in the community's credibility indicators in reducing people's sharing intention of fake news headlines \cite{yaqub2020effects}. Adding to prior work, our study develops a social media-like website to further explore people’s perceptual and behavioral responses to social media posts instead of using news headlines as stimuli, \finaledit{as was done in} many past studies \cite{pennycook2019lazy, yaqub2020effects, moravec2020appealing}. Understanding how users interact with social media posts is particularly valuable because misinformation embedded in users’ posts can be more rapidly spread due to ease of access to the platform. Therefore, one focus of our study is to compare the differing effects of algorithmic- and \finaledit{community}-based misinformation labels on partisan's perceptions of political social media posts. \jedit{Using COVID-19 social media posts}, we examined the following overarching questions: 1) Can misinformation labels reduce partisans' perceived accuracy and believability of fake posts? 2) Do different types of misinformation labels (algorithm, \finaledit{community}, third-party fact-checker) exhibit different levels of influence on partisans' perceptions and sharing intentions of fake political posts? 3) How will partisans perceive different misinformation labels?

Consistent with prior work, our findings confirm that both algorithmic and third-party misinformation labels can reduce people's perceived accuracy and belief in \jedit{COVID-19 related} fake posts regardless of source ideology. In terms of \finaledit{community} labels, we find that it reduced only liberal participants’ beliefs in \jedit{COVID-19 related} fake posts regardless of source ideology but not for conservative participants. For conservative participants, \finaledit{community} labels reduced the believability of fake liberal posts but were not effective for fake conservative posts. However, algorithmic and third-party fact-checker indicators reduced conservatives' belief in fake conservative posts. One interesting finding of our study is that the algorithmic labels perform as well as third-party fact-checker labels in reducing partisans' belief in \jedit{COVID-19 related} fake posts.
\section{Related Work}

\subsection{Hyper-Partisan Misinformation}

Hyper-partisan misinformation refers to misleading information with strong partisan bias \cite{epstein2020will}. Hyper-partisan misinformation often portrays itself as functionally indistinguishable from `real news' but is more often a mix of genres, combining news, entertainment, and politically-charged opinions \cite{mourao2019fake}. Misinformation entered the national spotlight during the 2016 US Federal Elections, and in its aftermath, misinformation has gained even more attention during the COVID-19 pandemic \cite{pennycook2021shifting}. \aledit{An internal study conducted by Twitter discovered that political content with a conservative bias is routinely favored by Twitter's algorithm \cite{chowdhury_belli_2021}.} The pandemic has become an important, yet divisive, political issue, with 60\% of surveyed voters indicating it was `the most important' or `an important' factor in the 2020 Presidential election, and only 38\% of the ‘most important’ group voting for Trump \cite{cnn_2020}. Studies also find that belief in COVID-19 misinformation is highly correlated with distrust in science \cite{agley2021misinformation}. 

Past misinformation studies have explored more traditional political issues such as immigration \cite{hameleers2020misinformation, valdez2018believability}, \finaledit{former} President Trump \cite{clayton2020real}, \aledit{mandatory} vaccinations \cite{gesser2018correcting}, gun control \cite{gao2018label}, and abortion \cite{kirchner2020countering}. As the pandemic has become a global issue, it is essential to examine COVID-19 related hyper-partisan misinformation as COVID-19 also represents a unique political phenomenon. Furthermore, when analyzing health and vaccination-related issues, the tone and approach of interventions appear to be even more crucial and sensitive \cite{gesser2018correcting}.

A large body of literature has explored why people share hyper-partisan misinformation and how to reduce such information sharing. One widely acknowledged explanation is that partisans will preferentially trust news that is consistent with their existing political ideology regardless of its truthfulness \cite{epstein2020will}. Past work suggests that hyper-partisan misinformation is the context where such politically motivated reasoning is more likely to occur \cite{kahan2017misconceptions, epstein2020will}. In other words, partisans value political alignment more than veracity when sharing misinformation \cite{grinberg2019fake}. Thus, without intervention, partisans tend to believe the news that aligns with their beliefs regardless of its truth due to confirmation bias \cite{kim2019combating}. Another possible reason for this tendency is that moral and highly emotional framing of news stories also drives sharing of hyper-partisan misinformation over social media, engaging with readers' morality through selective framing \cite{xu2020drives}.

Some prior studies suggest that engaging in active reasoning and internal deliberation \cite{bago2020fake} or shifting attention to accuracy \cite{pennycook2021shifting} can decrease people’s belief in hyper-partisan false news and inadvertent misinformation sharing. Other studies find that a reduction of emotion intensity in news stories can result in less hyper-partisan misinformation sharing behavior \cite{xu2020drives}. Even so, not much work has explored how real-time interventions such as algorithmic and community-based labels can reduce partisans’ confirmation bias.

\subsection{Labelling Fake News} 

While some \finaledit{earlier} studies find fake news labels to be generally ineffective \cite{gao2018label,vraga2020creating, moravec2018fake}, a growing body of literature suggests the opposite \cite{yaqub2020effects, moravec2020appealing, bago2020fake}. Interventions such as fake news labels - including warnings, credibility indicators, and disclaimers - have been proven to be effective in reducing people's believability and sharing intention of fake news \cite{yaqub2020effects, moravec2020appealing, pennycook2020implied}.

Past work suggests that the presence of a \finaledit{misinformation} label on news headlines that align with users’ beliefs can trigger cognitive activity such as increased attention and increased time spent considering the headline \cite{moravec2018fake}, and thus may reduce people's beliefs in \finaledit{misinformation}. One recent work found that not only can a \finaledit{misinformation} label with a detailed warning message (`declared fake by 3rd party fact-checkers') significantly reduce people's believability of \finaledit{misinformation}, but a simple stop sign can also reduce people's believability of \finaledit{misinformation} by triggering people's gut reaction and natural intuition \cite{moravec2020appealing, dias2020emphasizing}.

Studying the proliferation of hyper-partisan news over social media is also an expanding field, with sharing intentions originating from politically polarizing new sources (such as Infowars or Breitbart) gaining increasing research focus. The sharing of news headlines from such low-credibility hyper-partisan news sites drives significant sharing activity, however, interventions such as a simple label can increase user discernment and reduce sharing behavior \cite{pennycook2021shifting}. Strong interventions (e.g., an intervention label\finaledit{l}ing potential misinformation as `rated false') have been shown to cause stronger effectiveness compared to weaker interventions (e.g., label\finaledit{l}ing potential misinformation as `disputed') in counter-acting belief in \finaledit{misinformation} \cite{clayton2020real}.Furthermore, labelling was found to be most effective when appealing to both the mind’s system 1 (automatic) as well as system 2 (deliberate) cognition, demanding the activation of the mind’s intuition as well as its rational thinking core \cite{moravec2020appealing}. 

Most prior work examined the effect of labels provided by third-party fact-checkers \cite{moravec2020appealing}. While the third-party fact-checking indicator is reliable, it often heavily relies on human moderation and can only verify news at a slow pace and on a small scale \cite{yaqub2020effects}. Algorithmic fact-checking is becoming an increasingly viable alternative with benefits such as a massively increased breadth, a reduction in potential bias, and increased applicability to alternative cultures and contexts.

Other studies also indicate that users are highly cognisant of the content they consume on social media and are sensitive to potential misinformation \cite{hameleers2020misinformation,seo2019trust,kirchner2020countering,kim2019combating,wang2020factors, avram2020exposure}. In these cases where labelling can affect users' perceptions of fake news though, the matter of labelling \finaledit{misinformation} is further complicated by a potential "implied truth" on \textit{unlabelled} news items that users may assume after seeing that some posts have accompanying labels indicating misinformation \cite{pennycook2020implied}. This implied truth effect could backfire if the correction of false beliefs results in increased misconceptions \cite{berinsky2017rumors}. One possible explanation for this phenomenon is the higher level of awareness and thoughtfulness that exists when performing in an active study that demands the users' attention, compared to the passive nature of consuming social media content.

A prior study has found that political beliefs, interest in politics, and education all heavily affect belief in news generally, however - surprisingly - believability did not vary significantly between true news and labelled false news \cite{valdez2018believability}. Furthermore, the researchers found that young, educated, and left-oriented users distrusted any news (true or false) on social media to a greater degree. In another study, researchers found that when labels indicating a news source’s ideology matched the user’s ideology, this significantly increase\finaledit{d} the user’s trust in news from that source \cite{gao2018label}. However, this study also discovered that, generally, adding credibility labels to articles - such as indicating that the news article was disputed - was not effective in combating \finaledit{misinformation} and its spread. Similarly, one study found that news confirming users’ political beliefs produced strong confirmation bias effects that were too large to overcome with labelling \cite{moravec2020appealing}. 

Many social media platforms have been active in educating their user bases on the potential perils of \finaledit{misinformation}, to desirous results \cite{thomas_2017}.
\anedit{However, as \citet{zannettou2021won} found, the levels and types of engagement with misinformation labeled posts on social media platforms has variations depending on partisan leaning. Notably, \citet{zannettou2021won} studied a collection of tweets and misinformation interventions primarily about the 2020 U.S. Presidential election, finding that Republican users were responsible for 72\% of re-shared tweets with misinformation interventions (compared to 11\% by Democrats). This study, however, examined real tweets which were limited to one type of label attributed to third-party fact-checkers. This indicates a need to study how alternative intervention labels on hyper-partisan misinformation may impact Republican users vs. Democrat users.} Because of the aforementioned benefits to alternative types of \finaledit{misinformation} detection, the need to investigate the effectiveness of those types is clear. Adapting previous literature written investigating the accuracy and believability of \finaledit{misinformation} in the context of multiple types of labelling systems is valuable because it could provide insight into the future of \finaledit{misinformation} detection.

\subsection{\finaledit{Community} vs. Algorithmic Misinformation Labels}

Given the limitations in scalability of third-party fact-checkers, researchers and companies continue to explore the potential and effectiveness of automated techniques for misinformation detection. One method currently being studied, notably being tested in the real world as well, is crowd-sourcing techniques for detecting \finaledit{misinformation} in real-time: Twitter is currently piloting a community-driven, real-time misinformation detection effort (Birdwatch\footnote{blog.twitter.com/en\_us/topics/product/2021/introducing-birdwatch-a-community-based-approach-to-misinformation.html}). A separate study has also investigated the accuracy of crowd-sourced \finaledit{misinformation} detection compared to third-party fact-checkers, finding that crowd-sourced detection can be as reliable and accurate \cite{allen2020scaling}. 

Another alternative to third-party fact-checkers in misinformation detection is misinformation detection algorithms. In recent years, researchers have had success in creating misinformation detection algorithms, using various machine learning methods to identify features and characteristics of \finaledit{misinformation} in order to perform things like identifying categories of \finaledit{misinformation} \cite{hosseinimotlagh2018unsupervised} and distinguishing between fake and real news article titles \cite{horne2017just} to promising results. As misinformation detection algorithms advance in their detection accuracy, it is valuable to understand how such algorithms would be received by users of a social media platform.

\citet{seo2019trust} investigated the effectiveness and trustworthiness of false news warnings based on its attribution source, comparing warnings ascribed to third-party fact-checkers websites vs. a machine learning (ML) algorithm. They conducted two experiments to assess the effect of the warnings on participants’ abilities to detect fake and true news and participants’ trust in warnings \cite{seo2019trust}. In their first experiment, while all warning types led to higher correct detection of \finaledit{misinformation}, only the fact-checking warning led to participants being able to detect more true news as well. In the second experiment, they enhanced the ML warning with more information about how the algorithm worked and removed source labels from the fact-checking warning. This enhanced ML warning resulted in the highest correct detection of both fake and true news. However, despite the efficacy of all warnings in increasing recognition of \finaledit{misinformation}, participants still held low trust in warnings \cite{seo2019trust}. This low trust may indicate that users require additional or different ways of understanding how an automated detection method works before they feel comfortable trusting it. Additionally, this study does not examine the differences between these labels and a misinformation label attributed to other public or community users.

%\citet{dietvorst2015algorithm}'s findings reveal an “algorithmic aversion” that echoes the apprehension seen towards algorithmic models. They conducted studies about human perception of algorithms used in forecasting scenarios, observing that although the model outperformed human predictions, participants tended to abandon the model more frequently after seeing it err than when seeing a human err. They explain this phenomenon may be attributed to an expectation of a model needing to be perfect before accepting it. In three of five of the studies, however, the participants held domain knowledge of the scenario and therefore may have been more confident in their own judgment. On social media platforms though, where topics are broad and users are not necessarily domain experts, they may be more likely to use other signals such as a label before making a decision.

Most other past work on misinformation labels has primarily explored how users react to warnings attributed to third-party fact-checkers but not algorithmic labels or other public or community-based labels \cite{clayton2020real, kirchner2020countering, pennycook2020implied}. However, one recent study did include testing user perceptions of various credibility indicators including a "Public" and AI credibility indicator: \citet{yaqub2020effects} looked into how label interventions could affect people’s tendency to share fake stories. The authors manipulated the source that disputed the stories, testing 1) fact-checkers, 2) news media, 3) public (\finaledit{``}a majority of Americans\finaledit{''}), and 4) AI against people’s willingness to share headlines. The most effective indicator in reducing the sharing of fake headlines was fact-checkers, while AI was the least effective. There was also no significant difference in effect between the label attributed to the public compared to the AI label \cite{yaqub2020effects}. It is possible that the public label was not as effective as it was described as being \finaledit{``}a majority of Americans\finaledit{''}, rather than a more specific group of people leading participants to view it as too general. The AI label may have also been viewed skeptically as the use of the term \finaledit{``}Artificial Intelligence\finaledit{''} in describing the indicator may not have been well-received by participants who did not know the term or were familiar with it but uncertain about how the algorithm worked. In their experiment, however, the participants had a high rate of recall failure which may have affected their findings in the efficacy of the labels that they selected.

Our study intends to further examine this comparison between third-party fact-checker labels and automated, real-time labels (i.e. algorithmic labels and \finaledit{community} labels), particularly in the context of social media posts. Previous studies, including \cite{yaqub2020effects} have only examined the effect of these labels when shown with news article headlines. We also contribute to literature investigating how the effectiveness of these labels differ \finaledit{dependent} on people's \finaledit{party affiliation} and the ideology of the posts they view. 

%\citet{clayton2020real} tested the effects of providing a general warning to readers and/or two types of fake news labels on headlines — one of the labels stated a headline was “Disputed by Snopes and Politifact” while the other warned “Rated False by Snopes and Politifact”. Both types of fake news labels, particularly the latter, led to a reduced belief in fake news, while the general warning was related to possible spillover effect and reduced belief in true headlines.
\jedit{\section{Research Questions and Hypotheses}
Our study intend\finaledit{ed} to examine the following overarching research questions: 

\jedit{
\begin{itemize}
    \item \textbf{RQ1} Can misinformation labels reduce partisans' perceived accuracy and believability of fake posts?
    
    \item \textbf{RQ2} Do different types of misinformation labels (algorithm, \finaledit{community}, third-party fact-checker) exhibit different levels of influence on partisans' perceptions and sharing intention of fake political posts?
    
    \item \textbf{RQ3} How will partisans perceive different misinformation labels?
    
\end{itemize}
}

Building on prior work \cite{yaqub2020effects, moravec2020appealing}, our study hypothesize\finaledit{d} that all three labels \finaledit{would} be effective in reducing people's perceived accuracy and believability of fake posts regardless of the source ideology:

\begin{itemize}
    \item \textbf{H1}: Algorithmic labels, \finaledit{community} labels, and third-party fact-checkers' labels will be effective in reducing people’s perceived accuracy and believability of fake posts that both align (\textbf{H1a}) and do not align (\textbf{H1b}) with their beliefs.
\end{itemize}
    %\item \textbf{H2}: Algorithmic labels (\textbf{H2a}), users' labels (\textbf{H2b}), and third-party fact-checkers' labels (\textbf{H2c}) will be effective in reducing people’s perceived accuracy and believability of fake posts that do not align with their beliefs.
}

Previous studies have examined whether machine vs. human sources of online information affect the perceived bias and credibility of news \cite{wang2021moderating, waddell2019can, tandoc2020man}. Those studies find initial evidence that people’s perceptions of news bias may be attenuated when news is attributed to a machine cue. Some scholars have named such a phenomenon as the \finaledit{``}machine heuristic\finaledit{''}, which refers to the mental shortcut where people tend to consider machines as being more mechanical, objective, and ideologically unbiased than humans \cite{sundar2019machine}. Users often attribute bias to other human users rather than algorithms because they believe the algorithm is less likely to have a political agenda \cite{myers2018censored}. %Adding to previous work, we predicted that: 

One recent study \finaledit{found} that when information challenges people's views, they tend to reject such information and perceive it as less credible and more biased, regardless of whether its source is AI or a human \cite{wojcieszak2021can}. In other words, in regards to cross-cutting messages that do not align with partisans' beliefs, AI does not perform better than humans \cite{wojcieszak2021can}. This can possibly be explained by the fact that news that challenges people's opinion usually receives little cognitive activity and cues such that any label, machine or human, will be more likely to be ignored \cite{moravec2018fake}. Thus, we predicted the following hypothesis:

\begin{itemize}
    \item \textbf{\jedit{H2a}}: For posts that align with their beliefs, the algorithmic \finaledit{misinformation} labels will be more effective in reducing people’s perceived accuracy of fake posts than \finaledit{community} \finaledit{misinformation} labels.
    
    \item \textbf{\jedit{H2b}}: For posts that do not align with their beliefs, \jedit{there will be no difference in effectiveness between algorithmic \finaledit{misinformation} labels and \finaledit{community} \finaledit{misinformation} labels in reducing people’s believability of fake posts.}
\end{itemize}

Based on previous research that suggests third-party fact-checkers' labels were the most effective among \finaledit{misinformation} labels \cite{yaqub2020effects}, we further predicted that:

\begin{itemize}
    \item \textbf{\jedit{H3}}: For posts that align with their beliefs, the third-party fact-checkers' labels will be more effective in reducing people’s perceived accuracy and believability of fake posts than algorithmic (\textbf{\jedit{H3a}}) and \finaledit{community} (\textbf{\jedit{H3b}}) labels.

    \item \textbf{\jedit{H4}}: For posts that do not align with their beliefs, the third-party fact-checkers' labels will be more effective in reducing people's perceived accuracy and believability of fake posts than algorithmic (\textbf{\jedit{H4a}}) and \finaledit{community} (\textbf{\jedit{H4b}}) labels.
\end{itemize}

Recent studies suggest that there was a dissociation between accuracy judgment and people's sharing intention \cite{pennycook2021shifting}. People may still share news that they do not necessarily believe in \cite{pennycook2021shifting}. \jedit{Therefore, we also intended to examine the impact of labels on people's sharing, liking, and commenting intentions and people's perceptions of labels as exploratory analysis:

\begin{itemize}
    \item How will different labels affect people's sharing, liking, and commenting intentions?
    \item Will people have different perceptions of algorithmic, \finaledit{community}, and third-party fact-checkers' labels?
\end{itemize}
}

\section{Method}

\subsection{Participants}
We recruited 2257 participants using MTurk Toolkit on CloudResearch, an online participant pool that aggregates multiple market research platforms \cite{litman2017turkprime}. Participants were all from the United States. Participants were required to have a HIT approval rate greater than 95\% and \anedit{be over} 18 years old. After ruling out people who had moderate political views ($n$ = 353), \anedit{were} under the age of 18 ($n$ = 3), failed the recall ($n$ = 151), failed the embedded attention check question ($n$ = 67), and spent less than four minutes ($n$ = 6), 1677 participants remained in the data analysis.

% \usepackage{graphicx}
\begin{table}[!h]
\centering
\resizebox{.7\textwidth}{!}{%
\begin{tabular}{llccc}
\toprule
                                    &                      & \multicolumn{2}{c}{\textbf{Source Ideology}} & \textbf{}      \\ \cmidrule{3-4}
                                    &                      & \textbf{Conservative}   & \textbf{Liberal}   & \textbf{Total} \\ \midrule
\textit{\textbf{Conservative Participant}} & \textbf{Algorithm}   & 112                     & 96                 & 208            \\
\textbf{}                           & \textbf{\finaledit{Community}}        & 92                      & 111                & 203            \\
\textbf{}                           & \textbf{Third-party} & 104                     & 97                 & 201            \\
\textbf{}                           & \textbf{No Label}    & 90                      & 107                & 197            \\ \cmidrule{2-5} 
\textit{\textbf{Liberal Participant}}      & \textbf{Algorithm}   & 106                     & 100                & 206            \\
\textbf{}                           & \textbf{\finaledit{Community}}        & 99                      & 121                & 220            \\
\textbf{}                           & \textbf{Third-party} & 115                     & 115                & 230            \\
\textbf{}                           & \textbf{No Label}    & 110                     & 102                & 212            \\ \cmidrule{2-5} 
\textbf{Total}                      & \textbf{}            & 828                     & 849                & 1677           \\ \bottomrule
\end{tabular}%
}
\vspace{0.05in}
\caption{Number of Participants in Each Condition}
\label{tab:condition}
\end{table}

\begin{table}[!h]
\centering
\resizebox{\textwidth}{!}{%
\begin{tabular}{@{}ll@{}}
\toprule
\textbf{Label}                        & \textbf{Description}                                                                                             \\ \midrule
\multirow{2}{*}{Algorithm}   & {\textbf{Label}}: You may want to know this post’s accuracy is disputed by a misinformation detection algorithm.   \\ \cmidrule(l){2-2} 
 &
  \begin{tabular}[c]{@{}l@{}}{\textbf{Hyper-Text}} : We worked with third-party fact-checkers to develop an algorithm. Our algorithm helps us \\ detect misinformation quickly and accurately.\end{tabular} \\ \midrule
\multirow{2}{*}{\finaledit{Community}}        & {\textbf{Label}}: You may want to know this post’s accuracy is disputed by other users checking misinformation.    \\ \cmidrule(l){2-2} 
 &
  \begin{tabular}[c]{@{}l@{}}{\textbf{Hyper-Text}} : We worked with our users to create a community- based system. Our designated users help\\ us detect misinformation quickly and accurately.\end{tabular} \\ \midrule
\multirow{2}{*}{Third-party} & {\textbf{Label}}: You may want to know this post’s accuracy is disputed by third-party fact-checkers.              \\ \cmidrule(l){2-2} 
                             & {\textbf{Hyper-Text}} : We worked with third-party fact-checkers. Our goal is to detect misinformation accurately. \\ \bottomrule
\end{tabular}%
}
\vspace{-0.06in}
\caption{Label and Hyper-Text Description}
\label{tab:labeldescription}
\end{table}

% -0.2in

The mean age of the participants was 40.24 years old (\textsl{SD}=12.22, \textsl{Median}=38). Among 1677 participants: 744 (44.4\%) were male, 915 (54.6\%) were female, and 18 people chose other categories. When asked to self-report their political leaning, 868 of the participants self-reported as liberal-leaning (52\%), and 809 (37\%) as conservative-leaning \footnote{More detailed demographic data - including splits based on participant political ideology - can be found in the Supplemental Materials}. Participants were compensated \$1.50 US dollars for completing the experiment (Median completion time = 12.5 minutes excluding participants that kept their browsers open over 30 minutes).

\subsection{Experiment Design}
We conducted a 4 (labels: algorithm, \finaledit{community}, third-party fact-checkers, no label) X 2 (source ideology: liberal vs. conservative) between-subjects design online experiment ($N$ = 1677). Each participant was randomly assigned to one of the 8 conditions and read 12 posts. \anedit{The} 6 true and 6 false posts \anedit{displayed} had their veracity verified by major fact-checking organizations and were used as stimuli. \jedit{The number of participants in each condition is shown in Table~\ref{tab:condition}.}

% \aledit{Previous studies have included presenting a variety of number of posts including 8 \cite{kim2019combating}, 9 \cite{clayton2020real}, 10 \cite{moravec2018fake} 12 \cite{moravec2020appealing}, 13 \cite{valdez2018believability}, 14 \cite{gao2018label}, 16 \cite{kirchner2020countering} 24 \cite{seo2019trust}, 30 \cite{pennycook2020fighting}, and 36 \cite{pennycook2021shifting}. The number of 12 was chosen because many of these studies focus exclusively on presenting news headlines, whereas the longer format of Tweets requires more active reading by participants, lengthening the experiment.}

\subsubsection{Manipulation Check}
Several statistical tests were conducted to check whether randomization was effective and successful. One-way ANOVA showed there were similar sample distributions in terms of age, race, gender, education, Twitter usage, COVID-19 knowledge, political leaning, and party affiliation across 8 conditions. 

Two manipulation checks were embedded in the experiment. Since the understanding of the label plays a primary role in our experiment, the first manipulation check used the recall question as a filter to filter out people who could not successfully recall the label ($n$ = 151). \jedit{Participants who correctly answered the recall question \finaledit{``}could you recall the labels under the posts you just read?\finaledit{''} remained in the data analysis.} The second manipulation check tested whether the manipulation of source ideology was effective. \jedit{Participants needed to answer the question \finaledit{``}what, if any, is the political bias of this post?\finaledit{''} on a 7-point scale with 1 representing `extremely liberal' and 7 representing `extremely conservative' (adapted from \cite{munson2010presenting}) to rate the perceived bias of each post.} Repeated measures ANOVA showed that our manipulation of source ideology was successful. \jedit{Repeated measures ANOVA was used because participants were repeatedly measured on the same dependent variables for 12 posts.} Participants who were assigned to the conservative source ideology conditions rated the perceived bias of posts ($M$= 5.07, $SE$=.02) significantly higher than those who were assigned to the liberal source ideology conditions ($M$= 3.45, $SE$=.02), $p$ < .001. 

\subsubsection{Label Design}
All the fake posts were correctly labelled as misinformation \jedit{based on the ground truth. Only the description of the entity that labeled the posts varied across the conditions (algorithmic, \finaledit{community}, or third-party fact-checkers' labels)}. This design decision was made to understand the effect of the label type without confounding factors that can come from differences in the posts that different label types may tag as misinformation. 

%We also included a training process to avoid the potential effects introduced by the order of true or fake posts. 

 %Although previous studies found that direct and strong warnings were more effective than indirect warnings \cite{moravec2020appealing}, other research suggests that an "implied truth" effect may be produced from the use of strong warning labels because people may believe false unlabelled information more, dulling the label's overall effects \cite{pennycook2020implied}.%

The label designs were inspired by real-life social media fact check labels. The official misinformation policies of Twitter\footnote{blog.twitter.com/en\_us/topics/product/2021/introducing-birdwatch-a-community-based-approach-to-misinformation.html\label{birdwatch}}, Facebook\footnote{www.facebook.com/business/help/2593586717571940}, and Instagram\footnote{about.instagram.com/blog/announcements/combatting-misinformation-on-instagram} were instrumental, as the labels contained many phrases lifted directly from the policy statements. In addition to the direct labelling of the post as potentially misleading, participants also had access to hover-able hypertext that expands on how the treatments are intended to function in a real-world context. \jedit{Detailed label and hyper-text descriptions are shown in Table~\ref{tab:labeldescription}}.

\anedit{To create our label wording, we carefully reviewed labels of past studies and real labels used by social media companies for misinformation. Past studies such as \cite{moravec2020appealing, seo2019trust, kirchner2020countering, pennycook2018prior} characterized misinformation as being ``disputed'' by third-party fact-checkers. Social media companies such as Twitter and Facebook have used stronger language in declaring misinformation as verified by independent fact-checkers or experts such as adding an additional step and message before people can read a post labeled as misinformation (Twitter\footnote{https://blog.twitter.com/en\_us/topics/product/2020/updating-our-approach-to-misleading-information}) and blurring content with the message "False Information"and labeling content with the message ``The primary claims in the information are factually inaccurate'' (Facebook\footnote{https://techcrunch.com/2020/03/03/trump-coronavirus-hoax-fact-check/}). Because of the imperfect nature and potential biases of misinformation detection by algorithms or \finaledit{community}, we chose to use subtle language to nudge the reader to consider the presence of misinformation, telling them that they \finaledit{``}\textit{may want to know} this post's accuracy is \textit{disputed} by (source)'' as opposed to language that declares a post to be false}\aledit{\footnote{Although Birdwatch was officially announced by Twitter in January of 2021, many of the specific details regarding the UI were not released outside the private beta until June/July, and remains constantly updating. Because this study began collecting data in March 2021, the specific UI and wording of Birdwatch was unknown to the researchers until after the study was designed.}.}

\subsubsection{Social Media Website Development}
\begin{figure*}[!t]
	\centering
	%\vspace{-0.15in}
	\mbox{
		\subfigure[Website Interface
		\label{neu_lib}]{\includegraphics[width=0.49\textwidth]{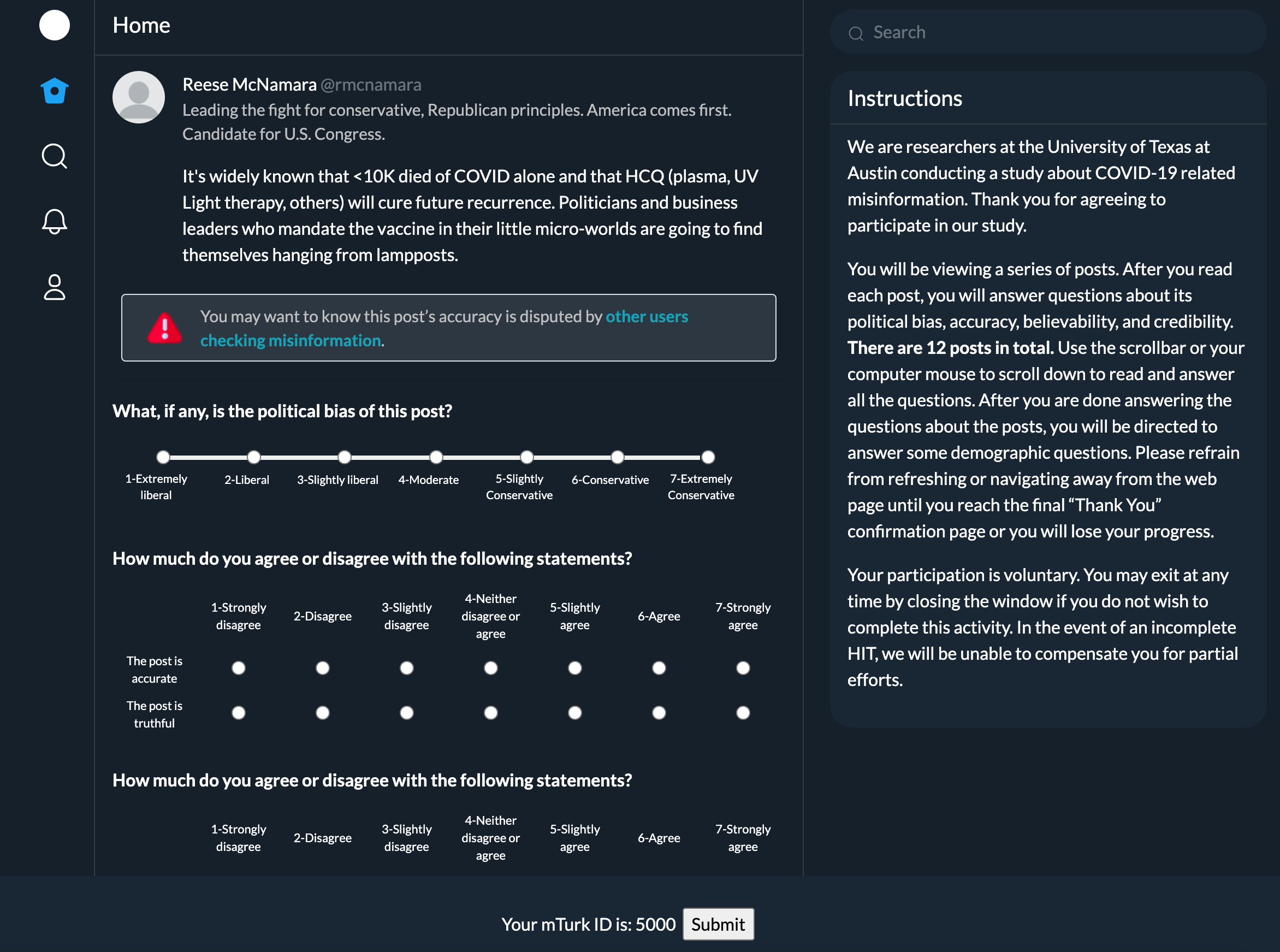}}
		
		\subfigure[Tutorial and Attention Check Interface \label{rev_lib}]{\includegraphics[width=0.49\textwidth]{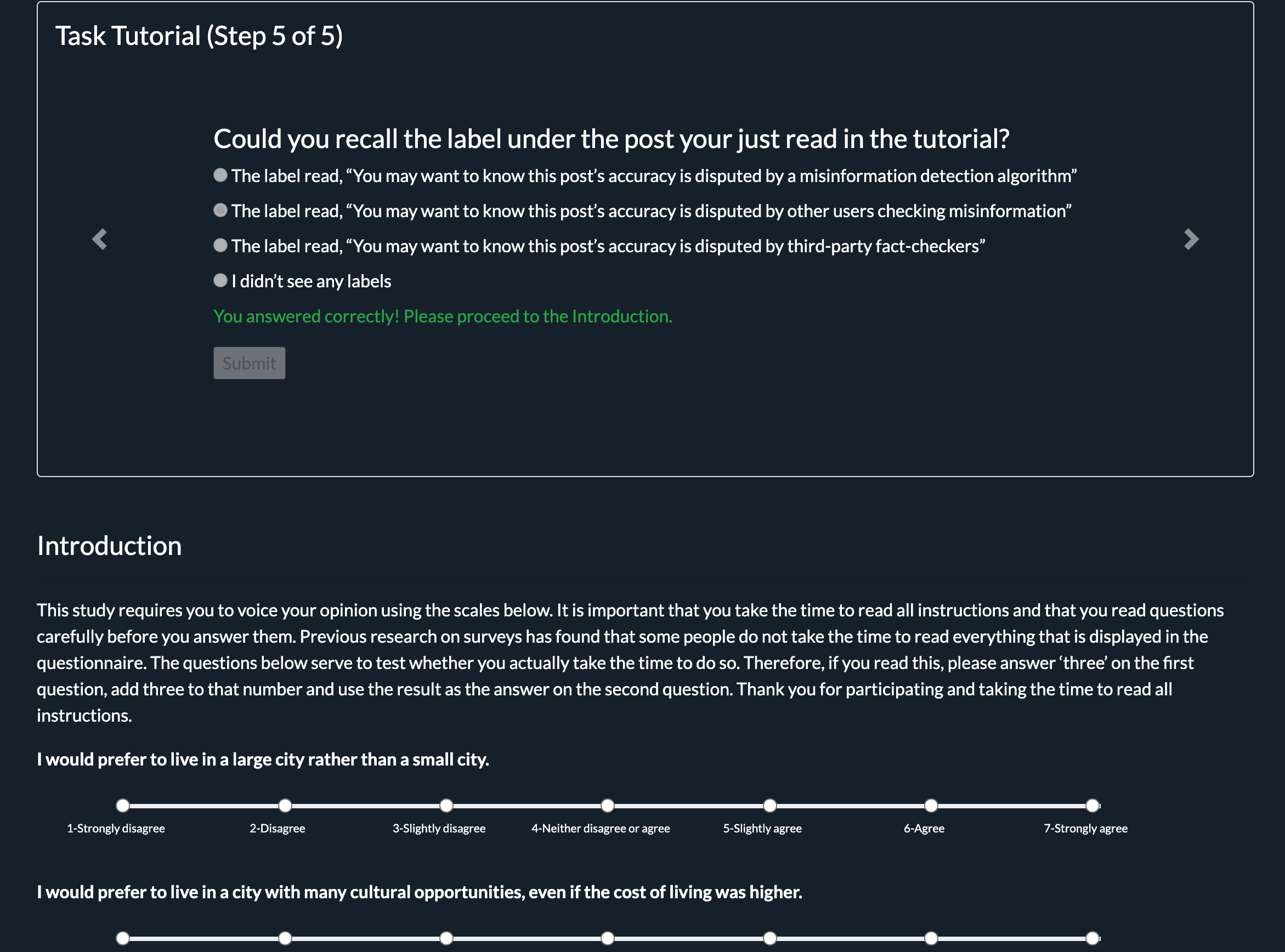}}
		}
	\vspace{-0.15in}
	\caption{Website and Tutorial Interface} 
	\label{fig:interface}
\end{figure*}

In order to simulate the real social media consumption environment, participants were asked to participate in the experiment on a website developed by our researchers. We created a website that emulates the Twitter environment. The website was developed in Python and hosted on Heroku, a cloud application platform; the data was recorded in a Heroku Postgres database. \jedit{The interface of the website is shown in ~ Figure \ref{fig:interface}}.

\anedit{\finaledit{Adding to previous experimental work on Twitter's misinformation labels \cite{zannettou2021won}, we chose to simulate Twitter as our social media platform because Twitter has taken multiple approaches to identify misinformation including third-party and community-driven fact-checking\footnotemark[4].} \finaledit{Additionally, most} past studies have studied misinformation within the context of \finaledit{Facebook} news headlines, while our work adds an additional layer of understanding by investigating misinformation labels in the context of another \finaledit{news format}, social media posts on Twitter. \finaledit{Finally}, we believe that simulating Twitter allows our study to have practical implications for companies including Twitter which has previously employed third\finaledit{-}party fact-checkers for misinformation labels and has recently been exploring platform \finaledit{community}-based fact-checking}.
%^above footnotemark is whatever the birdwatch url is

% Adding to previous experimental work on Twitter's misinformation labels \cite{zannettou2021won},
% .

\subsection{Procedures}

Once the participants reached the landing page of our experiment website, they were randomly assigned to one of the 8 conditions. They began with a tutorial walking them through an example of the task ahead. The tutorial specifically included a slide that explained some posts included a misinformation detection label (in non-control conditions) to ensure the participants would be aware of the label and its attribution (by an algorithm, third-party fact-checkers, or other platform users) during the experiment. To ensure the participant actively read the tutorial, we concluded the tutorial with a question asking them to recall what source the label was attributed to. Participants could not proceed to the main experiment until they answered this attention check question correctly, and participants were allowed multiple attempts. A satisfactory manipulation recall rate (91.00\%) was achieved, after concluding the tutorial.

After passing the tutorial and two attention check questions, participants proceeded to the main experiment where they were exposed to 12 social media posts. During the main experiment, participants were asked to evaluate a series of questions related to post believability and accuracy, as well as intention to share, comment, or like. Following the main experiment, participants completed a post-test questionnaire, where they were asked to evaluate the label effectiveness, objectiveness, whether the label was objective or politically unbiased, and if they belonged to one of the three label conditions. They were also asked a series of questions regarding their demographics, COVID-19 concern, and beliefs on machine learning and social media.

\subsection{Stimuli}

Stimuli were collected directly from actual posts on Twitter.com. \jedit{We collected 6 true and 6 false COVID-19 related social media posts. Both true and fake COVID-19 related posts were verified to be true or fake using third-party fact-checkers such as Snopes.com and PolitiFact, or large reputable health organizations such as the CDC or WHO. Each post was edited to have a similar length and readability score tested by the Flesch-Kincaid Grade Level ($M$ = 11.10, $SD$ = 1.64).} The topics of these posts vary from vaccines, treatments, mortality rate, and masks.

\jedit{Previous studies have presented different numbers of stimuli for participants to review \cite{kim2019combating, clayton2020real, moravec2018fake, moravec2020appealing, valdez2018believability, gao2018label, kirchner2020countering, seo2019trust, pennycook2020fighting, pennycook2021shifting}. For instance, \citet{kim2019combating} use 8 stimuli in their study; \citet{moravec2018fake} use 10 stimuli; \citet{moravec2020appealing} and \citet{kim2019says} use 12 news headlines; and \citet{seo2019trust} use 24 stimuli. The number of 12 stimuli was consistent with many prior studies \cite{moravec2020appealing, kim2019says}. We did not include more stimuli because many prior studies focus exclusively on presenting news headlines, whereas the longer format of social media posts in our study requires more active reading by participants, lengthening the experiment\jedit{\footnote{Specifically, the average number of characters of 12 social media posts in our stimuli is 193.42. The average number of characters of 12 news headlines used by \citet{moravec2020appealing} and \citet{kim2019says} is 76.08.}}.}

% have included presenting a variety of number of posts including 8 \cite{kim2019combating}, 9 \cite{clayton2020real}, 10 \cite{moravec2018fake}, 12 \cite{moravec2020appealing}, 13 \cite{valdez2018believability}, 14 \cite{gao2018label}, 16 \cite{kirchner2020countering} 24 \cite{seo2019trust}, 30 \cite{pennycook2020fighting}, and 36 \cite{pennycook2021shifting}.

%alternative? Previous studies have presented a wide range of posts for users to review, from between 8 to 16 posts \cite{kim2019combating, clayton2020real, moravec2018fake, moravec2020appealing, valdez2018believability, gao2018label, kirchner2020countering} to 24 to 36 \cite{seo2019trust, pennycook2020fighting, pennycook2021shifting}.

Most political-related Twitter experimental studies use elites/politicians or media organizations as the account types \cite{van2019priming}. Substantial evidence suggests that elite messages can transfer to public consciousness and conversation \cite{druckman2001implications}. Adapted from previous studies \cite{van2019priming}, our study uses political candidate bios to be our source to indicate the posts' ideology. We chose politically neutral posts and then manipulated the source ideology. For each post, we included a bio indicating the party affiliation and identity of the candidate (e.g., `Morgan Lang: Liberal. Let's leave the Earth a better place than we found it. Candidate for U.S. Congress.'; `Blake Walls: Conservative. No taxation without Representation. No big government! Candidate for U.S. Congress'). In order to reduce the impact of news source-specific effect, we followed previous research \cite{kim2019combating} and randomly assigned gender-neutral names to political candidates. All the names and bios were randomly assigned to posts for each participant. \jedit{Consistent with prior work \cite{kim2019combating, valdez2018believability}, we used default profile pictures to avoid potential gender or race bias}. %\aledit{similar to other comparable studies \cite{kim2019combating, valdez2018believability}}

Profile biographies were standardized to a similar Flesch-Kincaid Grade Level Score (Conservative: $M$ = 5.48, $SD$ = 2.19; Liberal: $M$ = 5.31, $SD$ = 1.70). The bios themselves consisted first of a direct description of the user’s intended political affiliation (liberal/conservative), followed by a statement or two of coded language related to their affiliation to increase the platform immersion. For example, a liberal user may indicate that ‘Transgender Lives Matter’ in their bio. After that, the biography indicated that the user was a candidate for US Congress; this was chosen to simulate a national-level politician with a realistic large social media reach, in contrast to a local-level politician without such reach.

\subsection{Pre-Test of the Source Ideology Manipulation}
We manipulated the profile bio of neutral social media posts to indicate the post ideology. In order to make sure the manipulation of the source ideology was successful, we conducted a pre-test ($N$=82) to test whether the direction of ideology was what we expected. Participants \footnote{Detailed demographic data regarding the pre-test participants can be found in the Supplemental Materials} were either randomly assigned to a group where all the neutral posts were assigned to liberal bios or a group where all the posts were assigned to conservative bios. We asked participants to answer the question "what, if any, is the political bias of this post?" on a 7-point scale with 1 representing `extremely liberal' and 7 representing `extremely conservative' (adapted from \cite{munson2010presenting}). Repeated measures ANOVA showed that our pre-test was successful. Posts assigned to conservative bios ($M$= 5.16, $SE$=.13) were rated significantly higher than those assigned to liberal bio ($M$= 3.50, $SE$=.15), $F$ (1, 80) = 50.65, $p$ < .001. 

\subsection{Measures}
\subsubsection{Post-level measures} After each post, we asked the following questions to understand participants' perception of the post.

\textbf{Political Bias}. Political bias was measured by asking participants, \finaledit{``}What, if any, is the political bias of this post?\finaledit{''} on a 7-point scale ranging from 1 indicating `extremely liberal' to 7 indicating `extremely conservative'.

\textbf{Perceived Accuracy}. Perceived accuracy was measured as an index consisting of two items by asking how much do participants agree or disagree with the following statements - \finaledit{``}the post is accurate\finaledit{''} and ``the post is truthful''  (adapted from \cite{pennycook2021shifting}) on a 7-point scales from 1 (Strongly disagree) to 7 (Strongly agree). The two items were highly correlated and can be averaged to form a reliable index (Cronbach’s $\alpha$ = .99).

\textbf{Perceived Believability}. Perceived believability was measured as an index consisting of two items by asking how much participants agree or disagree with the following statements - "the post is believable" and ``the post is credible'' on a 7-point scales from 1 (Strongly disagree) to 7 (Strongly agree), adapted from \cite{moravec2020appealing}. The two items were highly correlated and can be averaged to form a reliable index (Cronbach’s $\alpha$ = .96).

\textbf{Social Media Behaviors}. Participants were asked to evaluate the likelihood of them performing three typical social media actions: liking the post, sharing the post, and commenting on the post (adapted from \cite{moravec2020appealing}). Participants evaluated the likelihood on a 7-point scale ranging from 1 representing `extremely unlikely' to 7 representing \finaledit{`}extremely likely'.

\subsubsection{Post-survey measures}
In the post-survey, we asked the following questions to understand the participants' overall perceptions of the label. We also asked a recall-based manipulation check question.

\textbf{Label Perceptions}. In the post-test questionnaire, participants were asked to agree or disagree with three statements: `the label is mechanical', `the label is objective', and `the label is politically unbiased'. Participants rated each of the statements on a 7-point scale ranging from 1 indicating `strongly disagree' to 7 indicating `strongly agree'.

\textbf{Label Effectiveness}. In the post-test questionnaire, participants were asked, \finaledit{``}How effective is the label at indicating potential misinformation?\finaledit{''} and evaluated the label on a 7-point scale with 1 designating `extremely ineffective' and 7 designating `extremely effective' (adapted from \cite{moravec2020appealing}).

\textbf{Recall}. Participants were asked to recall the specific label condition they were assigned and were given five answer choices: one for each of the three label conditions, one for no labels, and one for if the participant did not remember.

Additional variable details and descriptions are available in the Supplemental Materials.
\section{Results}

We used linear mixed models to analyze data controlling for participant ID and post ID as random effects. Results showed that there was a significant 3-way interaction effect among the participant's political leaning, source ideology, and label, \textsl{F}(7,10046)= 6.26\jedit{\footnote{\jedit{We converted the data set from a wide format to a long format and obtained 10,062 data points for fake posts.}}}, $p$< .001 on the perceived accuracy of fake posts. There was a significant main effect of the labels, \textsl{F}(3,10046)=55.77, $p$< .001, and a significant main effect of \anedit{the} participant's political leaning, \textsl{F}(1,10046)=1328.22, $p$< .001, on the perceived accuracy of fake posts. 

\begin{figure*}[h]
	\centering
	%\vspace{-0.15in}
	\mbox{
		\subfigure[Perceived Accuracy (Conservative Participant) \label{neu_lib}]{\includegraphics[width=0.47\textwidth]{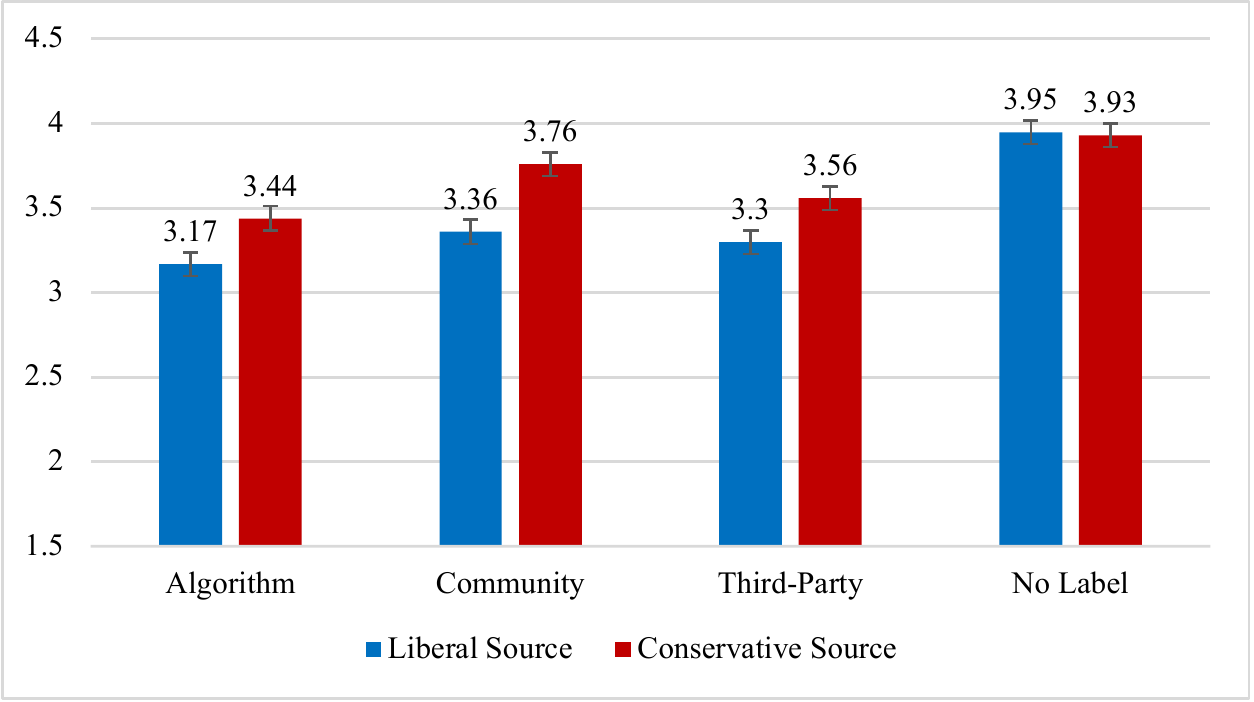}}
        
		\subfigure[Perceived Accuracy (Liberal Participant)\label{rev_lib}]{\includegraphics[width=0.47\textwidth]{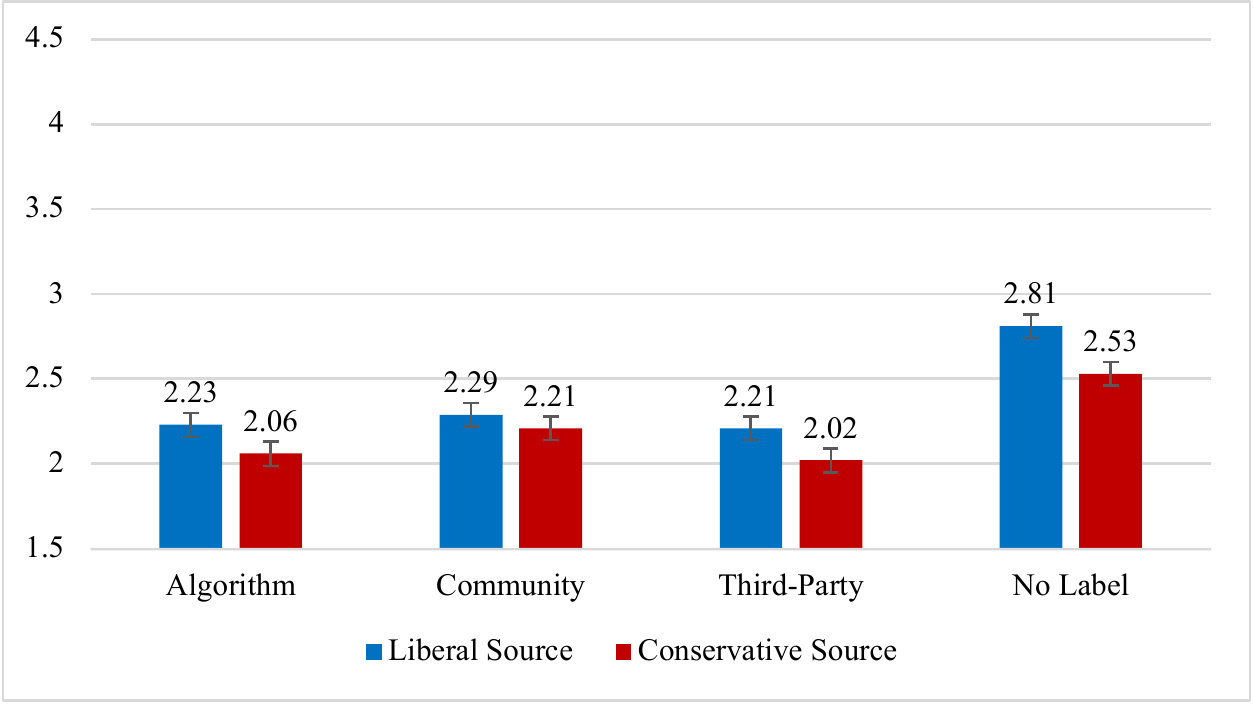}}
		}
	\\	
	\mbox{
		\subfigure[Perceived Believability (Conservative Participant) \label{neu_lib}]{\includegraphics[width=0.47\textwidth]{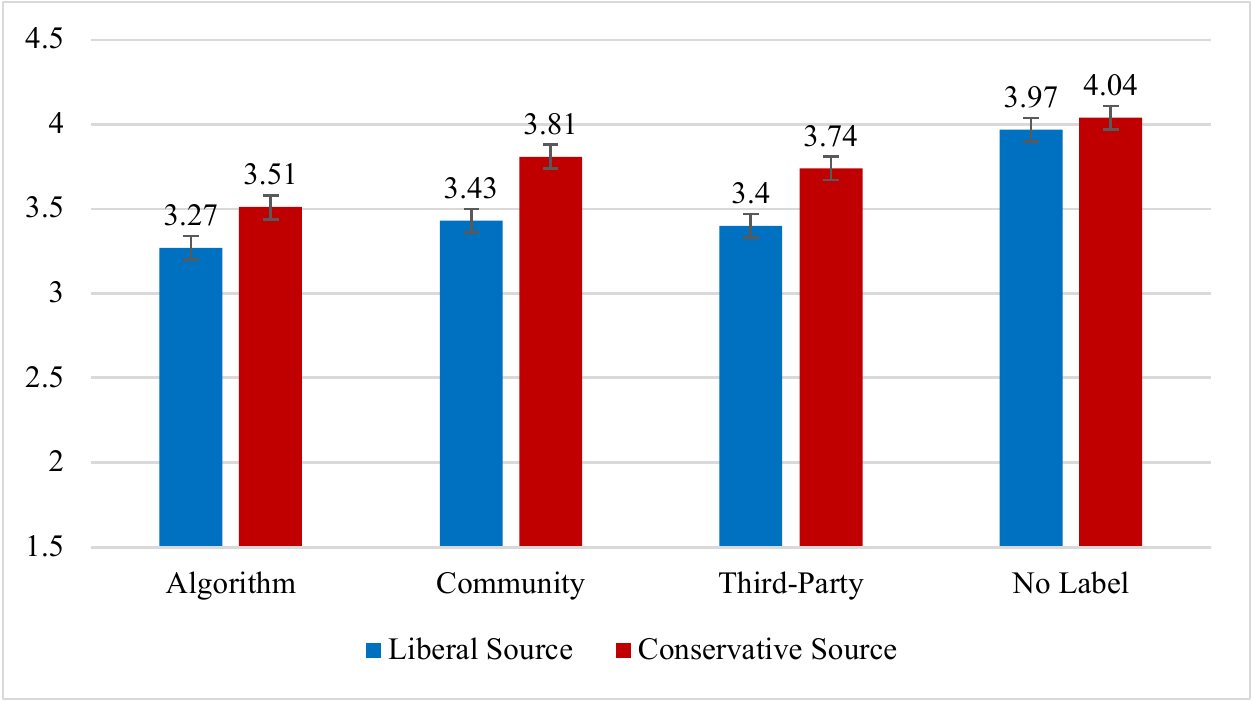}}
        
		\subfigure[Perceived Believability (Liberal Participant)\label{rev_lib}]{\includegraphics[width=0.47\textwidth]{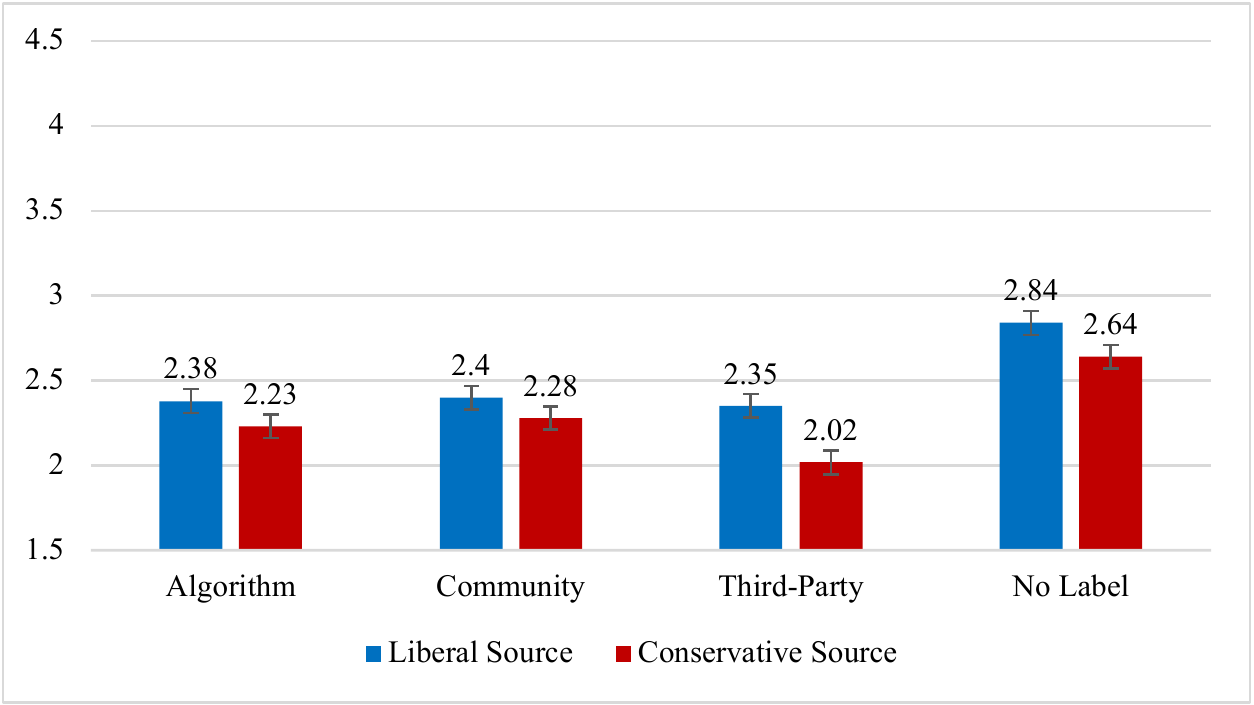}}
		}
	\vspace{-0.15in}
	\caption{The perceived accuracy and believability of fake posts with different labels} 
	\label{fig:accuracy}
\end{figure*}

\jedit{\textbf{H1} predicted that algorithmic labels, \finaledit{community} labels, and third-party fact-checkers' labels would be effective in reducing people’s perceived accuracy and believability of fake posts that both align (\textbf{H1a}) and do not align  (\textbf{H1b}) with their beliefs.} %\textbf{H1} predicted that algorithmic labels (\textbf{H1a}),  users' labels (\textbf{H1b}), and third-party fact-checkers' labels (\textbf{H1c}) would be effective in reducing people’s perceived accuracy and believability of fake posts that align with their political beliefs. 
\jedit{As shown in Figure~\ref{fig:accuracy} and Table ~\ref{tab:pairwise}}, multiple pairwise comparisons using the Bonferroni post-hoc test revealed that compared with the control (no label) condition, for both conservative- and liberal participants, algorithmic labels significantly reduced the perceived accuracy of fake posts that aligned with their political beliefs,  $p$< .001. For liberal participants, \finaledit{community} labels significantly reduced the perceived accuracy of fake posts that aligned with their beliefs, $p$< .001. For conservative participants, however, there was no significant difference between the  \finaledit{community} label condition (\textsl{M}=3.76, \textsl{SD}= .08) and the control (\textsl{M}=3.93, \textsl{SD}= .08) condition, $p$= .73 in terms of the perceived accuracy. Third-party fact-checkers' labels significantly reduced the perceived accuracy of ideologically agreeable fake posts for both conservative- and liberal partisans, $p$< .001. \jedit{Thus, (\textbf{H1a}) was partially supported. For fake posts that do not align with people's beliefs, multiple pairwise comparisons showed that all three labels significantly reduced the perceived accuracy of fake posts compared with the no label condition, $p$< .001. as shown in Figure~\ref{fig:accuracy} and Table ~\ref{tab:pairwise}. Thus, (\textbf{H1b}) was supported.} 

% \begin{figure}[!t]
% \centering
% \includegraphics[width=\textwidth]{figures/new_figure1.pdf}
% % \vspace{-0.25in}
% \caption{Caption here.}
% % \vspace{-2.5mm}
% \label{fig:overview}
% \end{figure}

% \usepackage{multirow}
% \usepackage{graphicx}
\begingroup
\setlength{\tabcolsep}{5pt}
\begin{table}[!t]
\centering
\resizebox{0.85\textwidth}{!}{%
\begin{tabular}{llcccc}
\toprule
\multirow{2}{*}{\textbf{\begin{tabular}[c]{@{}l@{}}Source \\ Ideology\end{tabular}}} &
  \multirow{2}{*}{\textbf{\begin{tabular}[c]{@{}l@{}}Label \\ Comparison\end{tabular}}} &
  \multicolumn{2}{c}{\textbf{Perceived Accuracy}} &
  \multicolumn{2}{c}{\textbf{Perceived Believability}} \\ \cmidrule{3-6} 
 &
   &
  \textit{\textbf{\begin{tabular}[c]{@{}c@{}}Mean\\ Difference\end{tabular}}} &
  \textit{\textbf{p}} &
  \textit{\textbf{\begin{tabular}[c]{@{}c@{}}Mean\\ Difference\end{tabular}}} &
  \textit{\textbf{p}} \\ \midrule
\multicolumn{6}{l}{\textit{\textbf{Conservative Participant}}}                      \\
Conservative & Algorithm / No Label    & -.49 & < .001*** & -.53 & < .001*** \\
             & \finaledit{Community} / No Label         & -.18 & .73       & -.23 & .27       \\
             & Third-Party  / No Label & -.37 & < .001*** & -.30 & < .001*** \\
             & Algorithm / \finaledit{Community}        & -.32 & .019*     & -.29 & .044*     \\
             & Algorithm / Third-Party & -.13 & 1.00      & -.22 & .21       \\
             & \finaledit{Community} / Third-Party      & .19  & .48       & .07  & 1.00      \\ \cmidrule{2-6} 
Liberal      & Algorithm / No Label    & -.78 & < .001*** & -.70 & < .001*** \\
             & \finaledit{Community} / No Label         & -.58 & < .001*** & -.55 & < .001*** \\
             & Third-Party  / No Label & -.65 & < .001*** & -.58 & < .001*** \\
             & Algorithm / \finaledit{Community}        & -.20 & .35       & -.16 & .86       \\
             & Algorithm / Third-Party & -.13 & 1.00      & -.13 & 1.00      \\
             & \finaledit{Community} / Third-Party      & .07  & 1.00      & .03  & 1.00      \\ \cmidrule{2-6} 
 &
   &
  \multicolumn{1}{l}{} &
  \multicolumn{1}{l}{} &
  \multicolumn{1}{l}{} &
  \multicolumn{1}{l}{} \\
\multicolumn{6}{l}{\textit{\textbf{Liberal Participant}}}                           \\
Conservative & Algorithm / No Label    & -.48 & < .001*** & -.41 & < .001*** \\
             & \finaledit{Community} / No Label         & -.32 & < .001*** & -.36 & < .001*** \\
             & Third-Party  / No Label & -.51 & < .001*** & -.51 & < .001*** \\
             & Algorithm / \finaledit{Community}        & -.15 & .54       & -.05 & 1.00      \\
             & Algorithm / Third-Party & .03  & 1.00      & .10  & 1.00      \\
             & \finaledit{Community} / Third-Party      & .08  & 1.00      & .14  & .71       \\ \cmidrule{2-6} 
Liberal      & Algorithm / No Label    & -.51 & < .001*** & -.46 & < .001*** \\
             & \finaledit{Community} / No Label         & -.51 & < .001*** & -.44 & < .001*** \\
             & Third-Party  / No Label & -.60 & < .001*** & -.49 & < .001*** \\
             & Algorithm / \finaledit{Community}        & .00  & 1.00      & -.02 & 1.00      \\
             & Algorithm / Third-Party & .03  & 1.00      & .03  & 1.00      \\
             & \finaledit{Community} / Third-Party      & .18  & .23       & .05  & 1.00      \\ \bottomrule
\end{tabular}%
}
\vspace{0.05in}
\caption{Pairwise comparisons of perceived accuracy and believability * $p<$ .05, ** $p<$ .01, *** $p<$ .001}
\label{tab:pairwise}
\vspace{-0.2in}
\end{table}
\endgroup

For perceived believability, the pattern stayed the same, \jedit{as shown in ~Figure \ref{fig:accuracy}}. There was a significant main effect of the labels, \textsl{F}(3,10046)=44.14, $p$< .001, and a significant main effect of participant's political leaning, \textsl{F}(1,10046)=1223.66, $p$< .001, on the perceived accuracy of fake posts. There was a significant 3-way interaction effect among the participant's political leaning, source ideology, and labels on the perceived believability, \textsl{F}(7,10046)= 6.42, $p$< .001. Pairwise comparisons showed that for both liberal and conservative participants, all three labels significantly reduced the perceived believability of fake posts that do not align with their political beliefs, $p$< .001. For fake posts that align with participants' political beliefs, both algorithmic and third-party fact-checkers' labels significantly reduced the perceived believability, $p$< .001; yet the significant effect only existed for liberal participants for the \finaledit{community} label condition, $p$< .001.

\jedit{\textbf{H2a}} predicted that for posts that align with their beliefs, the algorithmic labels would be more effective in reducing people’s perceived accuracy of fake posts than \finaledit{community} labels, which was partially supported for conservative participants. For conservatives, when the source ideology aligned with their political beliefs, algorithmic \finaledit{misinformation} labels (\textsl{M}=3.44, \textsl{SD}= .07) were more effective in reducing the perceived accuracy compared to \finaledit{community} labels (\textsl{M}=3.93, \textsl{SD}= .08), \jedit{$p$= .019}. For liberal participants, however, algorithmic \finaledit{misinformation} labels (\textsl{M}=2.297, \textsl{SD}= .07) were not significantly different from \finaledit{community} labels (\textsl{M}=2.293, \textsl{SD}= .06) when the posts aligned with their political beliefs.

\jedit{\textbf{H2b}} predicted that for posts that do not align with their beliefs, \jedit{there would be no difference in effectiveness between algorithmic \finaledit{misinformation} labels and \finaledit{community} \finaledit{misinformation} labels in reducing people’s believability of fake posts}, which \jedit{was} supported. For conservatives, when the source ideology does not align with their political beliefs, algorithmic labels (\textsl{M}=3.17, \textsl{SD}= .08) were not significantly different from \finaledit{community} labels in terms of perceived accuracy of fake posts (\textsl{M}=3.36, \textsl{SD}= .07), $p$= .35. For liberals, algorithmic labels (\textsl{M}=2.06, \textsl{SD}= .06) were not significantly different from \finaledit{community} labels in terms of perceived accuracy of fake posts (\textsl{M}=2.21, \textsl{SD}= .07), $p$= .54. 

\jedit{\textbf{H3}} predicted that for posts that align with their beliefs, the third-party fact-checkers' labels would be more effective in reducing people’s perceived accuracy and believability of fake posts than algorithmic \jedit{(\textbf{H3a}) and \finaledit{community} (\textbf{H3b}) labels}, which was not supported. Algorithmic labels performed as well as the third-party fact-checkers' labels \jedit{(\textbf{H3a})}, as shown in Table ~\ref{tab:pairwise}. Likewise, there was no significant difference in the effect of third-party fact-checkers' labels and \finaledit{community} labels \jedit{(\textbf{H3b})}.

% \usepackage{multirow}
% \usepackage{graphicx}
\begin{table}[!t]
\centering
\resizebox{1\textwidth}{!}{%
\begin{tabular}{llcccccc}
\toprule
\multirow{2}{*}{\textbf{\begin{tabular}[c]{@{}l@{}}Source Ideology\end{tabular}}} &
  \multirow{2}{*}{\textbf{Label}} &
  \multicolumn{2}{c}{\textbf{Sharing Intention}} &
  \multicolumn{2}{c}{\textbf{Liking Intention}} &
  \multicolumn{2}{c}{\textbf{Commenting Intention}} \\ \cmidrule{3-8} 
             &             & \textit{\textbf{Mean}} & \textit{\textbf{SE}} & \textit{\textbf{Mean}} & \textit{\textbf{SE}} & \textit{\textbf{Mean}} & \textit{\textbf{SE}} \\ \midrule
\multicolumn{6}{l}{\textit{\textbf{Conservative Participant}}}                                                                    &                        &                      \\
Conservative & Algorithm   & 2.43                   & .07                  & 2.17                   & .06                  & 2.37                   & .06                  \\
             & \finaledit{Community}        & 2.63                   & .08                  & 2.13                   & .07                  & 2.14                   & .07                  \\
             & Third-Party & 2.21                   & .07                  & 1.93                   & .06                  & 2.11                   & .07                  \\
             & No Label    & 2.65                   & .08                  & 2.31                   & .07                  & 2.34                   & .07                  \\ \cmidrule{2-8} 
Liberal      & Algorithm   & 2.08                   & .07                  & 1.76                   & .06                  & 1.92                   & .07                  \\
             & \finaledit{Community}        & 2.14                   & .07                  & 2.01                   & .06                  & 2.02                   & .06                  \\
             & Third-Party & 2.07                   & .07                  & 1.88                   & .06                  & 2.01                   & .07                  \\
             & No Label    & 2.26                   & .07                  & 2.01                   & .06                  & 2.11                   & .06                  \\ \cmidrule{2-8} 
             &             & \multicolumn{1}{l}{}   & \multicolumn{1}{l}{} & \multicolumn{1}{l}{}   & \multicolumn{1}{l}{} & \multicolumn{1}{l}{}   & \multicolumn{1}{l}{} \\
\multicolumn{6}{l}{\textit{\textbf{Liberal Participant}}}                                                                         & \multicolumn{1}{l}{}   & \multicolumn{1}{l}{} \\
Conservative & Algorithm   & 1.46                   & .05                  & 1.36                   & .04                  & 1.79                   & .06                  \\
             & \finaledit{Community}        & 1.56                   & .05                  & 1.46                   & .04                  & 1.94                   & .06                  \\
             & Third-Party & 1.41                   & .05                  & 1.37                   & .04                  & 2.03                   & .06                  \\
             & No Label    & 1.61                   & .05                  & 1.51                   & .04                  & 1.70                   & .06                  \\ \cmidrule{2-8} 
Liberal      & Algorithm   & 1.60                   & .05                  & 1.56                   & .05                  & 1.89                   & .06                  \\
             & \finaledit{Community}        & 1.65                   & .05                  & 1.51                   & .04                  & 1.96                   & .06                  \\
             & Third-Party & 1.68                   & .05                  & 1.50                   & .04                  & 1.83                   & .06                  \\
             & No Label    & 1.87                   & .06                  & 1.67                   & .05                  & 2.07                   & .06                  \\ \bottomrule
\end{tabular}%
}
\vspace{-0.07in}
\caption{The Means ($SE$) of sharing, liking, and commenting intention of fake posts}
\label{tab:share}
\end{table}

\jedit{\textbf{H4}} predicted that for posts that do not align with their beliefs, the third-party fact-checkers' labels would be more effective in reducing people’s perceived accuracy and believability of fake posts than algorithmic \jedit{(\textbf{H4a}) and \finaledit{community} (\textbf{H4b})} labels, which was not supported. There was no significant difference between the effect of third-party fact-checkers' labels and algorithmic labels \jedit{(\textbf{H4a})}. No significant difference was found between third-party fact-checkers' labels and \finaledit{community} labels \jedit{(\textbf{H4b})}.

\jedit{We conducted exploratory analyses to test} the effect of labels on people's sharing, liking, and commenting intention. We used linear mixed models while controlling for participant ID and post ID as random effects. Results showed that there was a significant main effect of labels on sharing, \textsl{F}(3,10046)=10.64, $p$< .001, and liking intention, \textsl{F}(3,10046)=13.37, but no main effect of labels on commenting intention, \textsl{F}(3,10046)= .85, $p$ = .47. There was a significant main effect of source ideology and participant's ideology on sharing, commenting, and liking intention, $p$< .001. There was also a significant 3-way interaction effect among the participant's political leaning, source ideology, and labels on sharing, liking, and commenting intention of fake posts, $p$< .001. We also reported the results of pairwise comparisons in Table ~\ref{tab:share} and Table~\ref{tab:share_pairwise} (in Appendix). The differences among label conditions were very small as people had overall low sharing, liking, and commenting intention.

Additional analyses were conducted to further examine the effect of labels on people's perceptions of \finaledit{misinformation} labels including label effectiveness, objectiveness, and whether the labels were mechanical and politically unbiased or not. Results showed that there were significant main effects of the label type, $p$< .001 and a significant main effect of participant's political leaning on people's label perceptions (effectiveness, politically unbiased, objectiveness, mechanical), $p$< .001. There was also a significant 3-way interaction effect among the participant's political leaning, source ideology, and label perceptions, $p$< .001. Pairwise comparisons showed that for both conservative and liberal participants, algorithmic labels were overall perceived as significantly more effective, politically unbiased, objective, and mechanical than \finaledit{community} labels for posts that both aligned and did not align with their beliefs. For conservative participants, algorithmic labels were even perceived as more politically unbiased, objective, and mechanical than third-party fact-checkers' labels. For liberal participants, however, the results were mixed. When liberal participants viewed conservative posts, they perceived third-party fact-checkers' labels as more politically unbiased, and more effective than algorithmic labels; when liberal participants viewed liberal posts, they considered algorithmic labels as more politically unbiased, as shown in Table~\ref{tab:fairness} and Table~\ref{tab:fairness_pairwise} (in Appendix).
\section{Discussion}

\subsection{Theoretical Implications}
Our study sheds light on the differing effects of various misinformation labels depending on partisan alignment. One interesting finding was that algorithmic misinformation labels outperformed \finaledit{community} misinformation labels when conservatives viewed ideologically consistent posts. One possible explanation for why algorithmic labels were more effective for hyper-partisan misinformation than \finaledit{community} labels is that people may have positive stereotypes about machine infallibility. Previous research suggests that the machine heuristic may reduce the perceived news bias \cite{wang2021moderating, waddell2019can}. Our results provided more evidence on the machine heuristic assumption \cite{sundar2019machine}. 

Another possible explanation for such results is conservative partisans’ mistrust of other social media users. Past work indicates the tendency of conservatives to be less trusting of media outlets, social media platforms, and journalists than liberals \cite{wessel2018messengers, vogels_perrin_anderson_2020, rauch2019comparing}. Between 1997 to 2016, trust in mainstream media by Democrats had decreased to 50\% while trust by Republicans had gone down from 41\% to 14\% \cite{wessel2018messengers}. Such mistrust by conservatives extends into perceptions of technology companies and social media platforms as well: 90\% of Republicans believe that social media platforms intentionally censor political opinions (vs. 59\% of Democrats), and 69\% of Republicans believe major technology companies favor liberal views over those of conservatives (vs. 19\% of Democrats believing the sites support conservative views over liberal ones) \cite{vogels_perrin_anderson_2020}. In a survey asking about attitudes towards journalists, conservative users on social media agreed more than liberals regarding statements that journalists were immoral, ego-driven, and affected biased mainstream media due to their own politics \cite{rauch2019comparing}. It is possible that the mistrust in social media platforms and journalists is projected onto our \finaledit{community} labels by conservative \finaledit{participants} who are wary of \textit{who} other users may be if selected by the platform. Conversely, they may view the algorithm as being a somewhat neutral source due to our explanation describing the algorithm as being created in conjunction with other platforms, which could make it appear more acceptable. 

Conservative users' skepticism towards social media platforms can also explain why conservative participants in our study rated all types of labels as less effective, objective, and politically unbiased than liberal participants. Conservatives' low trust in labels can be associated with another finding from our study that conservative participants rated fake posts with labels as more accurate and believable than liberal participants. \jedit{Our results also showed that conservative participants have overall higher perceived accuracy and believability of fake posts than liberal participants even for unlabeled posts. Such a result is not surprising as past work suggests conservatives are more likely to fall for \finaledit{misinformation} than liberals \cite{pennycook2020predictors}.}

Another interesting result indicated that \anedit{both algorithmic and third-party fact-checker indicators} reduced people's perceived accuracy and \finaledit{believability of} fake posts regardless of \finaledit{the post's} ideology. In fact, for conservative participants, algorithmic labels were even considered to be more politically unbiased, objective, and mechanical than third-party fact-checkers' labels. This drop in belief of third-party fact-checkers being unbiased reflects trends found in surveys conducted by Pew Research over attitudes held towards third-party fact-checkers. According to one Pew Research survey, 70\% of Republicans felt fact-checkers favored one side (while only 29\% of Democrats felt fact-checkers favored one side) \cite{walker_gottfried_2020}. Additionally, when asked their views on social media companies flagging inaccurate information, just 27\% of Republicans said they approved of this type of activity (vs. 73\% of Democrats that at least somewhat approve of this practice) \cite{vogels_perrin_anderson_2020}. Of the social media companies that have explained how their fact-checking process works, they typically outline a partnership with third-party fact-checking companies to verify potential misinformation \cite{allen2020scaling}, so it can be surmised that these results indicate mistrust of social media third-party fact-checkers already in place. Thus our results around the algorithmic label being seen as less biased than third-party fact-checkers combined with the algorithmic label reducing the accuracy of fake posts by conservative participants indicate a promising direction for additional work around the design of algorithmic labels. Many conservative users might be disapproving of labelling misinformation on social media platforms, but they may still pay attention to the labels if labels are attributed to sources that users deem acceptable or neutral.

Our work also broadens the field of misinformation studies by examining social media posts instead of news headlines. Past studies related to interventions with misinformation have primarily tested people's identification and rating of \finaledit{misinformation} through the format of news headlines. These studies present the headlines as they would appear on Facebook: in a standalone format with a photo, headline, and brief description \cite{clayton2020real, pennycook2019lazy}, and they may include a user's name and default profile photo that is intentionally blurred or nondescript \cite{kirchner2020countering,kim2019says}. In contrast, our study explores a new space: we build a website to simulate people’s perceptual and behavioral responses on social media and examine a new information format (user-level posts). We intentionally mimic Twitter, an environment where users are more likely to use functions such as "retweeting" to share articles and thus disseminate information much more rapidly compared to headlines on Facebook. Our web tool also displays names, account handles, and short bios to deepen a realistic social media platform experience. We include in the selection of posts displayed to participants some posts that have an insertion of personal beliefs or opinions (e.g., \textit{``It's widely known that <10k died of COVID alone and that HCQ (plasma, UV Light therapy, others) will cure future reoccurrence. Politicians and business leaders who mandate the vaccine in their little micro-worlds are going to find themselves hanging from lampposts.''}), which is atypical of headlines used in previous studies but common in social media posts. 

Our analyses reveal small but non-negligible effects of labels on people's sharing, liking, and commenting intention. The small differences among label conditions were not surprising as people had overall low sharing, liking, and commenting intention of those fake posts. This result was consistent with findings from past work. One recent study shows a discrepancy between people's sharing intention and accuracy judgment of fake posts \cite{pennycook2021shifting}. This may also explain why \citet{yaqub2020effects}'s study using sharing intent as the outcome variable exhibits some different patterns with our study. Their findings suggest that AI was least effective in reducing peoples' sharing intentions while ours suggest that algorithmic labels perform as well as third-party fact-checker labels and outperform \finaledit{community} labels in some cases. 

\subsection{Practical Implications}

Practically, we shed light on what types of misinformation labels are effective in nudging partisans to more mindfully assess and share social media content. Such results can have implications for widely used social media platform designs in the industry \anedit{about future mechanisms for misinformation detection and label design}. \anedit{Currently, fact-checking performed on social media websites is done by third-party fact-checkers which may be slow and fail to quickly identify harmful misinformation before it spreads, and recently, companies such as Twitter have begun to explore community-based fact-checking. Our study shows that misinformation identification attributed to algorithms or other platform users can have the same effect as third-party fact-checkers for improving a person's ability to accurately identify misinformation. This can allow for rapid detection of misinformation to prevent the spread of \finaledit{misinformation} more effectively than current techniques. 

However, the types of errors by real-time misinformation detection and people's perceptions of errors must be carefully assessed to gauge the real-world trust and impact of them before deploying these systems. Researchers have made strides in automated misinformation detection techniques such as creating larger and more robust data sets for algorithm training \cite{wang2017liar} and developing hybrid models that consider multiple dimensions of \finaledit{misinformation} such as content, metadata, source, and user engagement \cite{ruchansky2017csi}. However, achieving a highly accurate and accountable misinformation detection algorithm remains challenging \cite{shu2020fakenewsnet} due to the wide diversity of topics and features that cannot be encompassed in one data set \cite{shu2017fake} and the very nature of misinformation being to mislead and deceive users \cite{sharma2019combating}. Thus, although real-time detection methods offer the potential benefit of scaling to catch high volumes of misinformation ahead of harmful dissemination, they are also prone to error or biases which may contribute to harmful misinformation spreading or even decreased user trust in labels or platforms.}

The platform that we simulated, Twitter, has a very distinct user base in terms of both who uses the platform as well as who is actually actively producing Tweets and other forms of content. Twitter's users tend to be more Democratic than Republican, with the users producing the vast majority of Tweets also skewing Democratic \cite{policy_2021}. If aware of this information, the knowledge about users on Twitter could have also influenced conservative participants, and these participants may have been skeptical of the heterogeneity of users they imagined to be involved in labelling misinformation. \aledit{Republicans also believe that social media platforms intentionally censor political viewpoints they find objectionable at a much higher rate (90\%) compared to Democrats (59\%) \cite{vogels_perrin_anderson_2020}. 69\% of Republicans also believe that technology companies support the views of liberals over conservatives, compared to only 22\% that believe that they are supported equally \cite{vogels_perrin_anderson_2020}. These factors may result in a general distrust of community-based labelling among conservatives, believing that hand-selected users will skew liberal and be proxies for liberal-biased companies.} 

Less work exists exploring how users perceive misinformation in more informal, conversational settings such as Tweets on the Twitter platform where posts are often written and viewed with more fleeting attention and frequently integrate user opinion when discussing current events \cite{zhao2011comparing}. Additionally, while studying misinformation through how users can detect false news headlines is valuable, misinformation embedded in users’ posts can be more rapidly spread due to ease of access to the platform by almost everyone to post or share posts. Understanding misinformation in the context of forms like Tweets is of particular importance as users can and often do combine their own political opinions with misinformation, potentially exacerbating hyper-partisan tendencies and masking signals of misinformation. 

\aledit{Furthermore, prior studies found that ideological extremists are more likely to spread misinformation on Twitter compared to other social media platforms such as Facebook \cite{hopp2020people}. The reasoning behind this was that Facebook has an implied ‘real name’ policy providing a more normative social application built on personal and identity-linked information utilizing many of your real-world social connections. Twitter, on the other hand, is comparatively weak in rich personal interactions, resulting in the platform being rife with trolling and other anti-social behavior \cite{oz2018twitter}.}

%posts may reference current events but adopt tones such as sarcasm \cite{qazvinian2011rumor}

% Twitter was chosen as a platform to emulate for a couple of reasons to differentiate the study. 

%Twitter serves a demographically different user-base than Facebook, representing a generally smaller, younger, and more well-educated population \cite{auxier2021social}.

\aledit{Twitter's launching of its 'Birdwatch' pilot in January 2021 means that user- or community-based misinformation verification processes may become increasingly common.} Community-based platforms are potentially susceptible to abuse and misuse, and more needs to be studied regarding the potential for political exploitation, given social media's current role in American society. As the fight against misinformation develops, companies will be forced to adapt and create more innovative methods to confront users with the idea that their content may be misleading and betraying them. However, recent work has shown that crowd-sourcing for \finaledit{misinformation} detection does not necessarily result in the abuse of these reporting mechanisms, but can potentially produce reliable misinformation identification \cite{epstein2020will, allen2020scaling}. \jedit{Birdwatch relies on \finaledit{community-}based labeling whereas our study examines the effects of three types of labels (algorithmic, \finaledit{community}, and third-party labels).}

% Existing misinformation detection methods each have certain downsides - with algorithms being prone to errors and third-party experts requiring significant time - increasing the desire of social media companies to branch out and create alternative systems. However, the outcry over community-based systems already appears to be on the horizon. The potential for harassment and "brigading" is high, especially on a platform as polarized and anti-social as Twitter. Keith Coleman, Twitter's VP of Product, stated "We [Twitter] do expect to see brigading and things like that happening on Birdwatch," \cite{bond_2021}. %To add - how do our results change how people view Birdwatch? 

The results of \jedit{this paper} indicate that all three types of labels were generally effective in reducing the believability of fake posts, both for posts that align with the participant's political ideology, as well as for posts that don't. This is interesting because it demonstrates that the previously under-researched algorithmic and \finaledit{community-}based misinformation detection platforms are also effective in dissuading social media users. Our results confirmed that for conservative participants, algorithmic labels are more effective than \finaledit{community} labels in reducing the perceived accuracy of conservative posts. However, this is not the case for liberal participants, for whom \anedit{both \finaledit{community} labels and algorithmic labels were effective}. Despite this difference, the opportunity still opens for real-world social media platforms to install such \finaledit{community-}based misinformation detection algorithms. The advantages of \finaledit{community-}based (faster, more egalitarian, more transparent, larger-scale) and algorithmic (even faster, automated, highly manageable) techniques over more traditional third-party \finaledit{misinformation} detection methods are obvious. Research has indicated that crowd-based misinformation detection can even be as accurate as third-party independent fact-checkers \cite{allen2020scaling}. With these results, real-world platforms \jedit{such as Birdwatch} should feel more comfortable with rolling out these features in the near future.

\subsection{Limitations}

Despite our theoretical and practical contributions, this work had certain limitations. First, our study exposes participants to 6 true posts and 6 false posts, a greater ratio of false posts than any real social media platform. Even though such a ratio is commonly used in previous misinformation studies, it can potentially make people overall more skeptical to those posts by exposing them to several fake posts at one time. \anedit{Additionally, although we selected our total post count (12) shown to participants based on the number of posts \jedit{used in prior misinformation studies}, we recognize that this is only a small amount of posts a social media user may be exposed to in a day.} \jedit{We did an analysis of the effectiveness of the labels on each fake post. The main effect of label condition was significant across each fake post except for one single post (\textit{``It's widely known that <10k died of COVID alone and that HCQ (plasma, UV Light therapy, others) will cure future reoccurrence. Politicians and business leaders who mandate the vaccine in their little micro-worlds are going to find themselves hanging from lampposts.''}). The interaction effect of the label condition and the participant’s political ideology (i.e.,``algorithmic labels were more effective than \finaledit{community} labels when conservative users view conservative posts'') was significant only when we tested it with all fake posts; over a single post, it was either marginal significant or not significant possibly due to the small sample size.}

\anedit{Second, the set of posts we displayed to participants was limited to one context, COVID-19. While examining misinformation in this context is novel due to the emerging nature of the pandemic and information shared around it, the topics on social media platforms are diverse and may not carry the same societal polarization as COVID-19. Thus, this selection of COVID-19 posts may affect the generalizability of our results beyond COVID-19 related misinformation.} \aledit{Conservatives and people that ingest a steady diet of conservative news are at a higher susceptibility to incorrectly believe COVID-19 misinformation \cite{teng2021characterizing, calvillo2020political}. This does, however, potentially limit the implications of this study, because of COVID-19's role as a highly politicized ideological issue in American society, meaning the findings may not be exportable to other issues \cite{grossman2020political}.}

\anedit{Third}, we asked about political bias under each post, which may reinforce people's impression of source ideology, potentially leading to a less natural mindset than the real-world social media setting. Even though our website closely simulated social media platforms, it still has functional limits such as the inability to simulate true sharing behavior. Future work can measure people's actual sharing behavior on similar simulated websites instead of self-reported sharing intentions.

\aledit{A fourth limitation is the default Twitter profile pictures we used in the study. While it may be emblematic of 'bot' accounts with low credibility, the default icon was chosen to reduce potential gender and racial bias with human images.} \jedit{We also acknowledge that the stimuli we use were restricted to the text format so the results may not be generalized to social media posts in other formats such as images.}

\anedit{Fifth, even though we found that labels attributed to algorithms are effective in reducing the perceived accuracy of fake posts, we admit that algorithmically generated labels run the risk of making errors, including biased errors against specific groups \cite{barocas2016big, angwin2016machine}. Thus, while algorithms may be faster in labeling misinformation than professional fact-checkers, any algorithmic system designed for misinformation detection must consider how to reduce harm in users from mistakes, recover from mistakes made, and ensure that the algorithm itself is not biased in its detection such that errors or failures to detect misinformation do not exacerbate harm on specific populations.}

\anedit{Finally}, our participants were recruited from CloudResearch and thus were not nationally representative \cite{chandler2019online}. Moreover, workers from crowd-sourcing platforms such as CloudResearch and Amazon Mechanical Turk are known to have relatively high information literacy and technical expertise than the general population \cite{yaqub2020effects}. Such sample characteristics may affect the generalizability of our results.
\section{Conclusion}

Our work added a new dimension to hyper-partisan misinformation studies by examining the impacts of algorithmic, \finaledit{community} vs. third-party fact-checkers' labels depending on \finaledit{people's political ideology}. Our results showed that \anedit{both algorithmic and third-party fact-checkers' labels can reduce people's perceived accuracy and believability \finaledit{of} fake posts regardless of \finaledit{the post's} ideology, with no significant difference}. We also found that algorithmic labels were more effective in reducing people’s believability of fake posts than \finaledit{community} labels when conservative \finaledit{users} view ideologically-consistent posts. Our work sheds light on the effectiveness of real-time labels, which provides important theoretical and practical implications for automated and community-based misinformation detection approaches.

\section*{Acknowledgement}
We thank our participants who provided valuable insights. Our research was supported by UT Austin School of Information and Good Systems\footnote{https://goodsystems.utexas.edu}, a UT Austin Grand Challenge to develop responsible AI technologies.

\bibliographystyle{ACM-Reference-Format}
\bibliography{main_bib}

\section*{Appendix}

% \usepackage{multirow}
% \usepackage{graphicx}
\begin{table}[!h]
\centering
\resizebox{0.77\textwidth}{!}{%
\begin{tabular}{llcccc}
\toprule
\multirow{2}{*}{\textbf{Source Ideology}} &
  \multirow{2}{*}{\textbf{Label}} &
  \multicolumn{2}{c}{\textbf{Perceived Accuracy}} &
  \multicolumn{2}{c}{\textbf{Perceived Believability}} \\ \cmidrule{3-6} 
             &             & \textit{\textbf{Mean}} & \textit{\textbf{SE}} & \textit{\textbf{Mean}} & \textit{\textbf{SE}} \\ \midrule
\multicolumn{6}{l}{\textit{\textbf{Conservative Participant}}}                                                                    \\
Conservative & Algorithm   & 3.44                   & .07                  & 3.51                   & .07                  \\
             & \finaledit{Community}        & 3.76                   & .08                  & 3.81                   & .08                  \\
             & Third-Party & 3.56                   & .08                  & 3.74                   & .08                  \\
             & No Label    & 3.93                   & .08                  & 4.04                   & .08                  \\ \cmidrule{2-6} 
Liberal      & Algorithm   & 3.17                   & .08                  & 3.27                   & .08                  \\
             & \finaledit{Community}        & 3.36                   & .07                  & 3.43                   & .07                  \\
             & Third-Party & 3.30                   & .08                  & 3.40                   & .08                  \\
             & No Label    & 3.95                   & .07                  & 3.97                   & .07                  \\ \cmidrule{2-6} 
             &             & \multicolumn{1}{l}{}   & \multicolumn{1}{l}{} & \multicolumn{1}{l}{}   & \multicolumn{1}{l}{} \\
\multicolumn{6}{l}{\textit{\textbf{Liberal Participant}}}                                                                         \\
Conservative & Algorithm   & 2.06                   & .06                  & 2.23                   & .07                  \\
             & \finaledit{Community}        & 2.21                   & .07                  & 2.28                   & .07                  \\
             & Third-Party & 2.03                   & .06                  & 2.14                   & .06                  \\
             & No Label    & 2.53                   & .06                  & 2.64                   & .06                  \\ \cmidrule{2-6} 
Liberal      & Algorithm   & 2.30                   & .07                  & 2.38                   & .07                  \\
             & \finaledit{Community}        & 2.29                   & .06                  & 2.40                   & .06                  \\
             & Third-Party & 2.21                   & .06                  & 2.35                   & .06                  \\
             & No Label    & 2.81                   & .07                  & 2.84                   & .07                  \\ \bottomrule
\end{tabular}%
}
\vspace{0.05in}
\caption{The Means ($SE$) of perceived accuracy and believability of fake posts}
\label{tab:accuracy}
\vspace{-0.3in}
\end{table}
\begin{table}[!h]
\centering
\resizebox{1\textwidth}{!}{%
\begin{tabular}{llcccccc}
\toprule
\multirow{2}{*}{\textbf{\begin{tabular}[c]{@{}l@{}}Source \\ Ideology\end{tabular}}} &
  \multirow{2}{*}{\textbf{\begin{tabular}[c]{@{}l@{}}Label \\ Comparison\end{tabular}}} &
  \multicolumn{2}{c}{\textbf{Sharing Intention}} &
  \multicolumn{2}{c}{\textbf{Liking Intention}} &
  \multicolumn{2}{c}{\textbf{Commenting Intention}} \\ \cmidrule{3-8} 
 &
   &
  \textbf{\begin{tabular}[c]{@{}c@{}}Mean \\ Diff.\end{tabular}} &
  \textit{\textbf{p}} &
  \textbf{\begin{tabular}[c]{@{}c@{}}Mean \\ Diff.\end{tabular}} &
  \textit{\textbf{p}} &
  \textbf{\begin{tabular}[c]{@{}c@{}}Mean \\ Diff.\end{tabular}} &
  \textit{\textbf{p}} \\ \midrule
\multicolumn{6}{l}{\textit{\textbf{Conservative Participant}}}                         & \multicolumn{1}{l}{} & \multicolumn{1}{l}{} \\
Conservative & Algorithm / No Label    & -.22  & .24       & -.14   & .86       & .03                  & 1.00                 \\
             & \finaledit{Community} / No Label         & -.03  & 1.00      & -.17   & .45       & -.20                 & .28                  \\
             & Third-Party  / No Label & -.44  & < .001*** & -.371* & < .001*** & -.23                 & .10                  \\
             & Algorithm / \finaledit{Community}        & -.20  & .40       & .04    & 1.00      & .22                  & .10                  \\
             & Algorithm / Third-Party & .22   & .20       & .24    & .05*      & .26                  & .03*                 \\
             & \finaledit{Community} / Third-Party      & .41   & .00       & .20    & .21       & .04                  & 1.00                 \\ \cmidrule{2-8} 
Liberal      & Algorithm / No Label    & -.18  & .39       & -.24   & .03*      & -.18                 & .26                  \\
             & \finaledit{Community} / No Label         & -.12  & 1.00      & .00    & 1.00      & -.08                 & 1.00                 \\
             & Third-Party  / No Label & -.19  & .35       & -.12   & .99       & -.10                 & 1.00                 \\
             & Algorithm / \finaledit{Community}        & -.06  & 1.00      & -.24   & .03*      & -.10                 & 1.00                 \\
             & Algorithm / Third-Party & .00   & 1.00      & -.12   & 1.00      & -.09                 & 1.00                 \\
             & \finaledit{Community} / Third-Party      & .07   & 1.00      & .12    & .98       & .01                  & 1.00                 \\ \cmidrule{2-8} 
 &
   &
  \multicolumn{1}{l}{} &
  \multicolumn{1}{l}{} &
  \multicolumn{1}{l}{} &
  \multicolumn{1}{l}{} &
  \multicolumn{1}{l}{} &
  \multicolumn{1}{l}{} \\
\multicolumn{6}{l}{\textit{\textbf{Liberal Participant}}}                              & \multicolumn{1}{l}{} & \multicolumn{1}{l}{} \\
Conservative & Algorithm / No Label    & -.14  & .16       & -.16   & .03*      & .09                  & 1.00                 \\
             & \finaledit{Community} / No Label         & -.05  & 1.00      & -.06   & 1.00      & .24                  & .03*                 \\
             & Third-Party  / No Label & -.20  & .01**     & -.15   & .05*      & .33                  & < .001***            \\
             & Algorithm / \finaledit{Community}        & -.10  & .85       & .06    & .55       & -.16                 & .43                  \\
             & Algorithm / Third-Party & .05   & 1.00      & -.01   & 1.00      & -.24                 & .02*                 \\
             & \finaledit{Community} / Third-Party      & .15   & .13       & .09    & .80       & -.09                 & 1.00                 \\ \cmidrule{2-8} 
Liberal      & Algorithm / No Label    & -.27  & < .001*** & -.11   & .64       & -.18                 & .21                  \\
             & \finaledit{Community} / No Label         & -.23* & .01**     & -.16   & .07†      & -.11                 & 1.00                 \\
             & Third-Party  / No Label & -.20  & .05*      & -.17   & .05*      & -.24                 & .02*                 \\
             & Algorithm / \finaledit{Community}        & -.04  & 1.00      & .05    & 1.00      & -.07                 & 1.00                 \\
             & Algorithm / Third-Party & -.08  & 1.00      & .06    & 1.00      & .06                  & 1.00                 \\
             & \finaledit{Community} / Third-Party      & -.03  & 1.00      & .01    & 1.00      & .13                  & .58                  \\ \bottomrule
\end{tabular}%
}
\vspace{0.05in}
\caption{Pairwise comparisons of sharing, liking, and commenting intention † $p<$ .10, * $p<$ .05, ** $p<$ .01, *** $p<$ .001}
\label{tab:share_pairwise}
\end{table}
% \usepackage{multirow}
% \usepackage{graphicx}
\begin{table}[!t]
\centering
\resizebox{0.8\textwidth}{!}{%
\begin{tabular}{llcccccccc}
\toprule
\multirow{2}{*}{\textbf{Source Ideology}} &
  \multirow{2}{*}{\textbf{Label}} &
  \multicolumn{2}{c}{\textbf{Effective}} &
  \multicolumn{2}{c}{\textbf{\begin{tabular}[c]{@{}c@{}}Politically \\ Unbiased\end{tabular}}} &
  \multicolumn{2}{c}{\textbf{Objective}} &
  \multicolumn{2}{c}{\textbf{Mechanical}} \\ \cmidrule{3-10} 
 &
   &
  \textit{\textbf{Mean}} &
  \textit{\textbf{SE}} &
  \textit{\textbf{Mean}} &
  \textit{\textbf{SE}} &
  \textit{\textbf{Mean}} &
  \textit{\textbf{SE}} &
  \textit{\textbf{Mean}} &
  \textit{\textbf{SE}} \\ \midrule
\multicolumn{10}{l}{\textit{\textbf{Conservative Participant}}} \\
Conservative &
  Algorithm &
  4.35 &
  .07 &
  3.76 &
  .07 &
  4.32 &
  .06 &
  4.86 &
  .06 \\
 &
  \finaledit{Community} &
  3.78 &
  .07 &
  3.22 &
  .07 &
  3.64 &
  .07 &
  4.24 &
  .06 \\
 &
  Third-Party &
  4.10 &
  .07 &
  3.20 &
  .07 &
  3.89 &
  .06 &
  4.22 &
  .06 \\ \cmidrule{2-10} 
Liberal &
  Algorithm &
  4.19 &
  .07 &
  3.73 &
  .07 &
  4.20 &
  .07 &
  4.98 &
  .06 \\
 &
  \finaledit{Community} &
  3.91 &
  .07 &
  3.32 &
  .07 &
  3.94 &
  .06 &
  4.15 &
  .06 \\
 &
  Third-Party &
  4.30 &
  .07 &
  3.49 &
  .07 &
  3.80 &
  .07 &
  4.31 &
  .06 \\ \cmidrule{2-10} 
 &
   &
  \multicolumn{1}{l}{} &
  \multicolumn{1}{l}{} &
  \multicolumn{1}{l}{} &
  \multicolumn{1}{l}{} &
  \multicolumn{1}{l}{} &
  \multicolumn{1}{l}{} &
  \multicolumn{1}{l}{} &
  \multicolumn{1}{l}{} \\
\multicolumn{10}{l}{\textit{\textbf{Liberal Participant}}} \\
Conservative &
  Algorithm &
  5.46 &
  .04 &
  4.92 &
  .06 &
  5.09 &
  .06 &
  5.03 &
  .06 \\
 &
  \finaledit{Community} &
  5.17 &
  .04 &
  4.69 &
  .06 &
  4.74 &
  .06 &
  4.28 &
  .06 \\
 &
  Third-Party &
  5.67 &
  .04 &
  5.20 &
  .06 &
  5.21 &
  .05 &
  4.38 &
  .06 \\ \cmidrule{2-10} 
Liberal &
  Algorithm &
  5.39 &
  .05 &
  5.28 &
  .06 &
  5.05 &
  .06 &
  4.73 &
  .06 \\
 &
  \finaledit{Community} &
  5.07 &
  .05 &
  4.62 &
  .06 &
  4.65 &
  .05 &
  3.94 &
  .06 \\
 &
  Third-Party &
  5.24 &
  .05 &
  4.99 &
  .06 &
  4.88 &
  .05 &
  4.65 &
  .06 \\ \bottomrule
\end{tabular}%
}
\vspace{0.05in}
\caption{The Means ($SE$) of label perceptions}
\label{tab:fairness}
\end{table}

\begin{table}[!h]
\centering
\resizebox{1\textwidth}{!}{%
\begin{tabular}{llcccccccc}
\toprule
\multirow{2}{*}{\textbf{\begin{tabular}[c]{@{}l@{}}Source \\ Ideology\end{tabular}}} &
  \multirow{2}{*}{\textbf{\begin{tabular}[c]{@{}l@{}}Label \\ Comparison\end{tabular}}} &
  \multicolumn{2}{c}{\textbf{Effective}} &
  \multicolumn{2}{c}{\textbf{\begin{tabular}[c]{@{}c@{}}Politically \\ Unbiased\end{tabular}}} &
  \multicolumn{2}{c}{\textbf{Objective}} &
  \multicolumn{2}{c}{\textbf{Mechanical}} \\ \cmidrule{3-10} 
 &
   &
  \textbf{\begin{tabular}[c]{@{}c@{}}Mean\\ Diff.\end{tabular}} &
  \textit{\textbf{p}} &
  \textbf{\begin{tabular}[c]{@{}c@{}}Mean \\ Diff.\end{tabular}} &
  \textit{\textbf{p}} &
  \textbf{\begin{tabular}[c]{@{}c@{}}Mean \\ Diff.\end{tabular}} &
  \textit{\textbf{p}} &
  \textbf{\begin{tabular}[c]{@{}c@{}}Mean \\ Diff.\end{tabular}} &
  \textit{\textbf{p}} \\ \midrule
\multicolumn{10}{l}{\textit{\textbf{Conservative Participant}}}                                                           \\
Conservative & Algorithm / \finaledit{Community}        & .57  & < .001*** & .54  & < .001*** & .68  & < .001*** & .61  & < .001*** \\
             & Algorithm / Third-Party & .25  & .02*      & .56  & < .001*** & .42  & < .001*** & .64  & < .001*** \\
             & \finaledit{Community} / Third-Party      & -.31 & .01**     & .02  & 1.00      & -.25 & .02*      & .02  & 1.00      \\ \cmidrule{2-10} 
Liberal      & Algorithm / \finaledit{Community}        & .28  & .01**     & .41  & < .001*** & .26  & .01*      & .83  & < .001*** \\
             & Algorithm / Third-Party & -.11 & .79       & .25  & .05*      & .39  & < .001*** & .67  & < .001*** \\
             & \finaledit{Community} / Third-Party      & -.39 & < .001*** & -.17 & .26       & .13  & .43       & -.16 & .13       \\ \cmidrule{2-10} 
 &
   &
  \multicolumn{1}{l}{} &
  \multicolumn{1}{l}{} &
  \multicolumn{1}{l}{} &
  \multicolumn{1}{l}{} &
  \multicolumn{1}{l}{} &
  \multicolumn{1}{l}{} &
  \multicolumn{1}{l}{} &
  \multicolumn{1}{l}{} \\
\multicolumn{10}{l}{\textit{\textbf{Liberal Participant}}}                                                                \\
Conservative & Algorithm / \finaledit{Community}        & .29  & < .001*** & .23  & .03*      & .35  & < .001*** & .75  & < .001*** \\
             & Algorithm / Third-Party & -.21 & < .001*** & -.29 & < .001*** & -.12 & .31       & .65  & < .001*** \\
             & \finaledit{Community} / Third-Party      & -.50 & < .001*** & -.51 & < .001*** & -.47 & < .001*** & -.10 & .67       \\ \cmidrule{2-10} 
Liberal      & Algorithm / \finaledit{Community}        & .32  & < .001*** & .66  & < .001*** & .40  & < .001*** & .79  & < .001*** \\
             & Algorithm / Third-Party & .15  & .10       & .29  & < .001*** & .17  & .07†      & .08  & 1.00      \\
             & \finaledit{Community} / Third-Party      & -.18 & .02*      & -.37 & < .001*** & -.23 & .01**     & -.71 & < .001*** \\ \bottomrule
\end{tabular}%
}
\vspace{0.01in}
\caption{Pairwise comparisons of labels perceptions † $p<$ .10, * $p<$ .05, ** $p<$ .01, *** $p<$ .001}
\label{tab:fairness_pairwise}
\end{table}

\received{July 2021}
\received[revised]{November 2021}
\received[accepted]{February 2022}

\end{document}